\documentclass[aps,reprint,superscriptaddress,nofootinbib,floatfix,longbibliography,preprintnumbers]{revtex4-1}
 \usepackage[dvipsnames]{xcolor}
 
 \usepackage{amssymb,amsmath,amsfonts,bm,slashed,soul,ragged2e,graphicx,epstopdf,hyperref,array,comment,esint,siunitx,multirow,makecell,bm,esvect,caption,subcaption,ulem}
 \usepackage[utf8]{inputenc}

\usepackage{hyperref}

\def\aap{Astron.\ Astrophys.\ }
\def\apj{Astrophys.\ J.\ }

\def\mnras{Mon.\ Not.\ Roy.\ Astron.\ Soc.\ }

\usepackage{titlesec}
\usepackage{mathtools}
\usepackage{float}
\newcommand{\R}{\mathcal{R}}
\newcommand{\I}{\mathcal{I}}
\newcommand{\G}{\mathcal{G}}
\soulregister{\cite}7 % register \cite
\soulregister{\citep}7 % register \citep
\soulregister{\citet}7 % register \citet
\soulregister{\ref}7 % register \ref
\soulregister{\pageref}7 % register \pageref

\newcommand{\p}{\partial}
\newcommand{\dd}{\mathrm{d}}
\newcommand{\ee}{\mathrm{e}}
\newcommand{\PA}{\mathrm{PA}}

\newcommand{\overbar}[1]{\mkern 1mu\overline{\mkern-1mu#1\mkern-1mu}\mkern 1mu}

\allowdisplaybreaks
\begin{document}

\title{Forward Ray Tracing and Hot Spots in Kerr Spacetime}
\author{Lihang Zhou}
\email{lzhou2@caltech.edu}
\affiliation{Walter Burke Institute for Theoretical Physics, California Institute of Technology, Pasadena, California 91125, USA}
\affiliation{Department of Astronomy, School of Physics, Peking University, Beijing 100871, China}
\affiliation{Kavli Institute for Astronomy and Astrophysics, Peking University, Beijing 100871, China}
\affiliation{Center of Gravity, Niels Bohr Institute, Blegdamsvej 17, 2100 Copenhagen, Denmark}

\author{Zhen Zhong}
\email{zhen.zhong@tecnico.ulisboa.pt}
\affiliation{CENTRA, Departamento de F\'{\i}sica, Instituto Superior T\'ecnico -- IST, Universidade de Lisboa -- UL, Avenida Rovisco Pais 1, 1049 Lisboa, Portugal}

\author{Yifan Chen}
\email{yifan.chen@nbi.ku.dk}
\affiliation{Center of Gravity, Niels Bohr Institute, Blegdamsvej 17, 2100 Copenhagen, Denmark}

\author{Vitor Cardoso}
\email{vitor.cardoso@nbi.ku.dk}
\affiliation{Center of Gravity, Niels Bohr Institute, Blegdamsvej 17, 2100 Copenhagen, Denmark}
\affiliation{CENTRA, Departamento de F\'{\i}sica, Instituto Superior T\'ecnico -- IST, Universidade de Lisboa -- UL, Avenida Rovisco Pais 1, 1049 Lisboa, Portugal}

\begin{abstract}

Hotspots, often characterized as pointlike emissions, frequently appear near black holes with significantly enhanced luminosity compared to the surrounding accretion flow. Notably, such hotspots are regularly observed near the black hole at the center of the Milky Way. Light rays emitted from these sources follow complex trajectories around the black hole before reaching distinct locations on the observer's image plane. Precisely resolving both direct emissions and their higher-order images—despite the latter's intensity suppression—is essential for extracting detailed spacetime information, including the black hole's mass, spin, and inclination angle. To improve the accuracy and efficiency of hotspot modeling, we develop a forward ray tracing method based on the analytic integral solution of Kerr geodesics, leveraging conserved quantities. Our approach traces geodesics from a given emission point near the black hole to a distant observer, effectively capturing multiple images with a tailored parametrization scheme for root-finding. By perturbing these geodesics, we map finite-size emissions to distinct regions on the image plane, enabling the quantification of image shapes and amplification rates. This method not only enhances the identification of strongly lensed photons from black holes but also enables efficient spacetime tomography and hotspot localization, leveraging observations from the Event Horizon Telescope and its upcoming next-generation upgrades.

\end{abstract}

\date{\today}

\maketitle

\tableofcontents

%%%%%%%%%%%%%%%%%%%%%%%
\section{Introduction}
%%%%%%%%%%%%%%%%%%%%%%%

The recent breakthroughs by the Event Horizon Telescope (EHT) employing the very long baseline interferometry (VLBI) technique have inaugurated a novel perspective for examining black holes (BHs) at the horizon scale \cite{EventHorizonTelescope:2019dse,EventHorizonTelescope:2022wkp}. The imaging of supermassive BHs (SMBHs), specifically Sgr A$^*$ and M87$^*$, has unveiled details of spacetime and plasma structures in regions dominated by strong gravity. 

Within the framework of (vacuum) general relativity, the spacetime geometry of an astrophysical BH is uniquely determined by its mass and spin. 
The current technique for estimating the mass of SMBHs entails examining the diameter of a ring structure, anticipated to be the critical curve where photons enter bound orbits around the SMBH~\cite{1965SvA.....8..868P,1968ApJ...151..659A,Luminet:1979nyg,Falcke:1999pj,Gralla:2019xty}. This approach has led to mass measurements of $(6.5\pm 0.7)\times 10^9 M_\odot$ for M87$^*$ \cite{EventHorizonTelescope:2019ggy} and $4.0^{+1.1}_{-0.6}\times 10^6 M_\odot$ for Sgr A$^*$~\cite{EventHorizonTelescope:2022exc}, where $M_\odot$ denotes the mass of the sun. The precision of these mass estimates is largely limited by the angular resolution available, leading to uncertainties that exceed those obtained from stellar motion analyses~\cite{GRAVITY:2021xju,Do:2019txf}. Determining the spin of SMBHs remains a formidable challenge. The dimensions and shape of the photon ring exhibit only a slight dependence on the BH spin, given the present limits of angular resolution. Consequently, current observational data can only hint at a preference for high spins, inferred indirectly through the analysis of accretion flows and jet structures~\cite{Cruz-Osorio:2021cob,EventHorizonTelescope:2022urf}.

The photon ring on the image plane consists of photons that orbit the BH multiple times before converging on the observer's image plane. More precise methods for measuring BH mass and spin involve the detection of higher-order images from photons that orbit more times than those arriving directly, indexed by an integer $N$ denoting the number of half-orbits~\cite{Johannsen:2010ru,Gralla:2019xty,Johnson:2019ljv,Gralla:2020srx,Broderick:2021ohx,Paugnat:2022qzy,Broderick:2022tfu,Wang:2022mjo}. By comparing emissions from direct ($N=0$) photons with those of $N=1$ and $2$, it is possible to simultaneously measure the BH mass and spin. However, such procedure necessitates angular resolutions attainable only by extending baselines into space~\cite{Johnson:2019ljv,Gralla:2020srx,Andrianov:2022snn,Gurvits:2022wgm,Shlentsova:2024qzj}. Currently more feasible methods, not requiring significant enhancements in angular resolution, leverage the time variability of emission sources and seek their correlations in both the time domain and across different regions near the photon ring~\cite{Broderick:2005my,Moriyama:2015zfa,Saida:2016kpk,Moriyama:2019mhz,Tiede:2020jgo,Wong:2020ziu,Hadar:2020fda,Chesler:2020gtw,Hadar:2023kau}. These correlations stem from multiple emissions originating from the same source, creating echoes with separations as predicted by general relativity. The time delay is sensitive to the BH mass, while the position angle shift on the photon ring for photons completing one additional half-orbit is indicative of the spin~\cite{Gralla:2019drh,Hadar:2020fda}. 

Hotspots, characterized by transient and localized emissions that significantly overshadow their surrounding areas, represent a promising source for detecting light echoes~\cite{Broderick:2005my,Broderick:2005jj,Moriyama:2015zfa,Saida:2016kpk,Moriyama:2019mhz,Tiede:2020jgo,Wong:2020ziu,Chesler:2020gtw,2022arXiv220403715L} compared to the extended accretion flow~\cite{Cardenas-Avendano:2024sgy}. Their intense emissions and pronounced variability facilitate the clear distinction between both their direct emissions and the secondary echoes against the background of accretion flows or jets. In the case of Sgr A$^*$, flares from hotspots are observed on a daily basis and span the electromagnetic spectrum~\cite{Trippe:2006jy,Marrone:2007tc,Witzel:2020yrp}, including the millimeter band~\cite{Doeleman:2008qh,Fish:2008ji,Doeleman:2008xq,Johnson:2014msa,EventHorizonTelescope:2022ago,Wielgus:2022heh}, near-infrared~\cite{Genzel:2003as,Eckart:2006fc,Meyer:2006fd,Zamaninasab:2009df,Do:2019vob,GRAVITY:2023avo}, and x-rays~\cite{Baganoff:2001kw,Porquet:2003ic,Kusunose:2010xs,2017MNRAS.472.4422K,Haggard:2019mro,Andres:2021cpw}. These hotspots can be situated remarkably close to the BH, within a few gravitational radii~\cite{gravity2018,GRAVITY:2020lpa,GRAVITY:2020hwn,Wielgus:2022heh,GRAVITY:2023avo}. Magnetic reconnection within the plasma is a plausible mechanism for their formation, as indicated by general relativistic magnetohydrodynamics (GRMHD) simulations~\cite{Younsi:2015xna,Dexter:2020cuv,Ripperda:2020bpz,Porth:2020txf,Dexter:2020cuv,Ripperda:2020bpz,Chatterjee:2020wef,Ripperda:2021zpn,Vos:2023ska,xi2024revisiting}. Observations in astrometry, polarimetry, and light curves of the flares enable the reconstruction of hotspot motion, providing a unique opportunity to probe plasma dynamics and spacetime by exploring different regions around the BH~\cite{Tiede:2020jgo,GRAVITY:2020lpa,Cardoso:2021sip,GRAVITY:2020hwn,Wielgus:2022heh,Vos:2022yij,vonFellenberg:2023hit,Levis:2023tpb,GRAVITY:2023avo,Vincent:2023sbw,Huang:2024wpj,Kocherlakota:2024hyq,Antonopoulou:2024qco,Yfantis:2024eab}.

Extracting spacetime information from a hotspot, particularly through its echoes, which can unveil universal features independent of the plasma composition, necessitates simulating its emission within the accretion flow around the SMBH, as exemplified in Refs.~\cite{Broderick:2005my,Meyer:2006fd,Trippe:2006jy,Hamaus:2008yw,Zamaninasab:2009df,Tiede:2020jgo,GRAVITY:2020hwn,Matsumoto:2020wul,Dokuchaev:2020rye,GRAVITY:2020lpa,2022arXiv220403715L,Wielgus:2022heh,Vos:2022yij,Rosa:2022toh,Guo:2022ghl,Aimar:2023kzj,Vincent:2023sbw,Rosa:2023qcv,Najafi-Ziyazi:2023oil,Yfantis:2023wsp,Levis:2023tpb,Chen:2023knf,vonFellenberg:2023hit,Huang:2024wpj,Rosa:2024bqv,Chen:2024ilc,Kocherlakota:2024hyq,Antonopoulou:2024qco,Yfantis:2024eab}. 
This task predominantly employs backward ray tracing, a method that traces geodesics from the observer's image plane back toward the BH~\cite{Luminet:1979nyg}, and is a foundational technique in most ray tracing and radiative transfer packages~\cite{Dexter:2016cdk,Moscibrodzka:2017lcu,Bronzwaer:2018lde,Pihajoki:2018ihj,Younsi:2019iee,Bronzwaer:2020kle,White:2022paq,Aimar:2023vcs,EventHorizonTelescope:2023hqy,Huang:2024bar}. However, efficiently simulating hotspot echoes presents challenges. The implementation of a finite grid on the image plane complicates the precise targeting of pointlike sources. While increasing the emission volume may assist, the probability of capturing higher-order images exponentially decreases with each additional image order. Moreover, accurately reconstructing the spatial distribution of emissions within the hotspot is challenging, as only a limited number of geodesics intersect the emission volume. These challenges extend to identifying strongly lensed photons and precisely determining hotspot locations and BH parameters from observational data.

Unlike backward ray tracing, which traces light paths from the observer back to the source, forward ray tracing initiates geodesics at the source and follows them to a distant observer. The main challenge is determining the initial directions of geodesics that reach the observer. One approach is to solve the geodesic equations in integral form using conserved quantities~\cite{Carter:1968rr,Bardeen:1973tla,Dexter_2009,Lupsasca:2024wkp} (see Eqs.~(\ref{eq:integral form}) and (\ref{eq:path integrals}) below) and search for images on the conserved quantity plane. This method has been applied to both localized point sources (e.g., Refs.~\cite{Cunningham1973,Vazquez:2003zm}) and extended emitters (e.g., Refs.~\cite{Beckwith:2004ae,Gelles:2021kti}).

In this work, we refine this approach by implementing the recent analytic geodesic solutions from Refs.~\cite{Gralla:2019ceu,Gralla:2019drh}, which provide a complete classification of geodesic motion types and analytic expressions for all the integrals in Eqs.~(\ref{eq:path integrals}). This enables efficient and precise calculations of geodesic directions in all cases. Additionally, we introduce a tailored parametrization scheme for root-finding, facilitating the computation of higher-order images that exponentially converge to the critical curve. As a result, our method is more efficient, accurate, and comprehensive in characterizing images of pointlike sources—especially higher-order images—while remaining applicable to general source and observer locations, improving upon previous works that relied on numerical integration~\cite{Cunningham1973,Dolence:2009zz} or were restricted to specific source locations, such as being infinitely far from the BH in Ref.~\cite{Vazquez:2003zm}.

Additionally, by introducing perturbations to the geodesics~\cite{Cunningham1973,Ohanian1987}, we map a finite emission volume onto specific regions of the image plane, enabling the quantification of image shapes and amplification rates. We confirm that higher-order images undergo spatial compression perpendicular to the critical curve due to the photon ring's instability, while in the parallel direction, these images may either compress or expand depending on the specific location of the emission. 
We apply our forward ray tracing method to realistic observation scenarios involving highly time-variable hotspots near M87$^*$ or Sgr A$^*$, conducting analyses to determine the hotspot location and the BH parameters. This approach demonstrates significant promise for elucidating the BH spin with the next-generation EHT (ngEHT)~\cite{Emami:2022ydq,Johnson:2023ynn,Ayzenberg:2023hfw}.

The structure of this paper is as follows: Section \ref{sec:NGBR} offers an overview of null geodesics within Kerr spacetime and the backward ray tracing method. Section \ref{sec:FRLE} presents the forward ray tracing method and its application in generating both direct and higher-order images from a point source. In Sec.~\ref{sec:distortion}, a perturbative mapping method is presented, which translates emissions from a finite volume to an array of regions on the image plane, allowing for a systematic analysis of image distortion as the order increases. Section \ref{sec:tomography} explores the application of forward ray tracing to spacetime tomography and hotspot localization. Section \ref{sec:dis} summarizes our findings and discusses potential future directions for both theoretical and observational studies. Numerical code \texttt{KerrP2P}~\cite{KerrP2P} used in this study is available~\footnote{\texttt{KerrP2P}: \href{https://github.com/AuroraDysis/KerrP2P}{https://github.com/AuroraDysis/KerrP2P}}. Throughout this study, we adopt a system of geometrized units where the gravitational constant $G$ and the speed of light $c$ are both set to $G=c=1$.

%%%%%%%%%%%%%%%%%%%%%%%%%%%%%%%%%%%%%%%%%%%%%%%%%
\section{Null Geodesics in Kerr Spacetime and Backward Ray Tracing}
\label{sec:NGBR}
%%%%%%%%%%%%%%%%%%%%%%%%%%%%%%%%%%%%%%%%%%%%%%%%%

This section provides an overview of null geodesics in Kerr spacetime, focusing on the analytic integral formalism as detailed in Refs.~\cite{Gralla:2019ceu,Gralla:2019drh}. We explore the trajectories of light rays that reach a distant observer, examining their motion and detailing how they are governed by two conserved quantities in Kerr spacetime. These quantities are then mapped to specific coordinates on the image plane. We conclude this section with a brief discussion of the traditional backward ray tracing method used to calculate BH images, paving the way for the discussion on the forward ray tracing method in the following section.

%%%%%%%%%%%%%%%%%%%%%%%%%%%%%%%%%%%%%%%%%%%%%%
\subsection{Null Geodesics in Kerr Spacetime}
\label{subsec:Trajectory}
%%%%%%%%%%%%%%%%%%%%%%%%%%%%%%%%%%%%%%%%%%%%%%
We utilize the Boyer-Lindquist coordinates $x^{\mu} = (t, r, \theta, \phi)$ to describe Kerr spacetime. The spacetime line element is given by
\begin{equation}
\begin{aligned}
\dd s^2=&-\frac{\Delta}{\Sigma}(\dd t-a\sin^2\theta\dd\phi)^2+\frac{\Sigma}{\Delta}\dd r^2+\Sigma\dd\theta^2\\
&+\frac{\sin^2\theta}{\Sigma}\Big[(r^2+a^2)\dd\phi-a\dd t\Big]^2,
\end{aligned}
\label{eq:Kerr metric}
\end{equation}
where the spin parameter $a \equiv {J}/{M}$ is defined, with $J$ as the BH angular momentum and $M$ its mass. Furthermore,  $\Sigma \equiv r^2 + a^2\cos^2\theta$ and $\Delta \equiv r^2 - 2Mr + a^2$.

We explore null geodesics within this background, representing the trajectories of massless photons. These photons possess four-momentum $p^{\mu} \equiv {\dd x^{\mu}}/{\dd \sigma}$, where $\sigma$ is the affine parameter. The path of each photon is determined by two conserved quantities~\cite{Carter:1968rr,Gralla:2019ceu,Lupsasca:2024wkp}
\begin{equation}
    \lambda\equiv\frac{p_\phi}{-p_t}, \quad
    \eta\equiv\frac{p_\theta^2-\cos^2\theta(a^2p_t^2-p_\phi^2\csc^2\theta)}{-p_t}, \label{eq:def lambda eta}
\end{equation}
which correspond to the energy-rescaled angular momentum and the Carter constant, respectively. These quantities enable us to express the geodesics in a first-order differential form
\begin{subequations}
\label{eq:p}
    \begin{align}
        \frac{\Sigma}{E}\frac{\dd r}{\dd \sigma}&=\nu_r\sqrt{\mathcal{R}(r)},\label{eq:pr}\\
        \frac{\Sigma}{E}\frac{\dd \theta}{\dd \sigma}&=\nu_{\theta}\sqrt{\Theta(\theta)},\label{eq:ptheta}\\
        \frac{\Sigma}{E}\frac{\dd \phi}{\dd \sigma}&= \frac{a}{\Delta}(r^2+a^2-a\lambda)+\frac{\lambda}{\sin^2\theta}-a,\label{eq:pphi}\\
          \frac{\Sigma}{E}\frac{\dd t}{\dd \sigma}&=\frac{r^2+a^2}{\Delta}(r^2+a^2-a\lambda)+a(\lambda-a\sin^2\theta),\label{eq:pt}
    \end{align}
\end{subequations}
where the radial and polar angle potentials, $\mathcal{R}(r)$ and $\Theta(\theta)$, are defined as
\begin{align}
    \mathcal{R}(r)&\equiv(r^2+a^2-a\lambda)^2-\Delta\left[\eta+(\lambda-a)^2\right],\\
    \Theta(\theta)&\equiv\eta+a^2\cos^2\theta-\lambda^2\cot^2\theta\,.
\end{align}
The conditions $\mathcal{R}(r) = 0$ and $\Theta(\theta) = 0$ define the turning points, indicating the reversal of directions for the momenta $p^r$ and $p^\theta$, respectively. These directional changes in momenta are encoded by the sign variables $\nu_r \equiv \operatorname{sgn}(p^r)$ and $\nu_\theta \equiv \operatorname{sgn}(p^\theta)$, which take values $\pm 1$ and represent the directionality of the respective momenta.

Considering a light ray emanating from a source point at $x^{\mu}_s=(t_s, r_s, \theta_s, \phi_s)$ and ultimately arriving at $x^{\mu}_f=(t_f, r_f, \theta_f, \phi_f)$, we can express Eqs.~(\ref{eq:p}) in integral form as follows
\begin{subequations}
    \begin{align}
        I_r&=G_{\theta} \,\equiv\tau,\label{eq:inteq1} \\
        \phi_f-\phi_s&=I_{\phi}+\lambda G_{\phi},\label{eq:inteq2} \\
        t_f-t_s&=I_t +a^2 G_{t}, \label{eq:inteq3}
    \end{align}
\label{eq:integral form}%
\end{subequations}
 where $\tau$, referred to as the ``Mino time''~\cite{Mino:2003yg}, accumulates along the ray's path. The $I$- and $G$-integrals, which depend on $r_f$ and $\theta_f$, respectively, are defined as
\begin{subequations}
    \begin{align}
        I_r&\equiv\fint_{r_s}^{r_f}\frac{1}{\sqrt{\mathcal{R}(r)}}\nu_r\dd r, \label{eq:Ir}\\
        I_{\phi}&\equiv\fint_{r_s}^{r_f}\frac{a(2Mr-a\lambda)}{\Delta\sqrt{\R(r)}}\nu_r\dd r, \label{eq:Iphi}\\
        I_{t}&\equiv\fint_{r_s}^{r_f}\frac{r^2\Delta+2Mr(r^2+a^2-a\lambda)}{\Delta\sqrt{\R(r)}}\nu_r\dd r, \label{eq:It}\\
        G_{\theta}&\equiv\fint_{\theta_s}^{\theta_f}\frac{1}{\sqrt{\Theta(\theta)}}\nu_{\theta}\dd \theta, \label{eq:Gtheta}\\
        G_{\phi}&\equiv\fint_{\theta_s}^{\theta_f}\frac{\csc^2\theta}{\sqrt{\Theta(\theta)}}\nu_{\theta} \dd \theta, \label{eq:Gphi}\\
        G_{t}&\equiv\fint_{\theta_s}^{\theta_f}\frac{\cos^2\theta}{\sqrt{\Theta(\theta)}}\nu_{\theta} \dd \theta, \label{eq:Gt}
    \end{align}
    \label{eq:path integrals}%
\end{subequations}
Here, $\fint$ signifies path integration along the light ray's trajectory between $x_s^{\mu}$ and $x_f^{\mu}$. The momentum signs $\nu_r$ and $\nu_\theta$ change at radial and angular turning points to ensure that $\nu_r \dd r > 0$ and $\nu_\theta \dd\theta > 0$ throughout the entire path. Mino time $\tau$ facilitates controlling the variation of the endpoint $x_f^{\mu}$ along the geodesic, such that $x^{\mu}_f=x^{\mu}_f(\tau)$. A practical approach for calculating these integrals involves introducing antiderivatives $\I_i\ (i=r,\phi,t)$ and $\G_j\ (j=\theta,\phi,t)$, which are defined such that their derivatives match the integrands in Eq.~(\ref{eq:path integrals}), excluding the $\nu_r$ or $\nu_\theta$ factors~\cite{Gralla:2019ceu,Gralla:2019drh}. For instance,
\begin{equation}
    \frac{\dd \I_r}{\dd r}\equiv\frac{1}{\sqrt{\R(r)}},
\qquad
    \frac{\dd \G_{\theta}}{\dd \theta}\equiv\frac{1}{\sqrt{\Theta(\theta)}},
\end{equation}
with other antiderivatives being similarly defined.

Each light ray originating from $x_s^{\mu}$ is uniquely identified by its initial direction, characterized by two conserved quantities, $\lambda$ and $\eta$, in addition to the initial directional signs $\nu_r^s$ and $\nu_{\theta}^s$. The specific values of $\lambda$ and $\eta$ influence the geodesics' behavior, resulting in various types of angular and radial motions, as thoroughly examined in Refs.~\cite{Gralla:2019ceu,Gralla:2019drh}. In the following, we succinctly overview the distinct motion patterns in both $\theta$ and $r$ directions and outline the parameter space of our interest.

 \paragraph{Polar angular motion} 
 The motion in the $\theta$ direction can be categorized into two types. For $\eta>0$, we observe ``ordinary'' motion, where the geodesic oscillates between two angular turning points, $\theta_-<\pi/2$ and $\theta_+=\pi-\theta_-$, traversing the equatorial plane. These angular turning points are defined as~\cite{Gralla:2019ceu,Gralla:2019drh}
 \begin{equation}
     \theta_{\pm}=\arccos(\mp\sqrt{u_+}),
 \end{equation}
 where
 \begin{equation}
    u_{\pm} \equiv \Delta_{\theta}\pm\sqrt{\Delta_{\theta}^2+\frac{\eta}{a^2}},\quad\Delta_{\theta} \equiv \frac{1}{2}\left(1-\frac{\eta+\lambda^2}{a^2}\right).
 \end{equation}

Conversely, for $\eta<0$, we encounter ``vortical''-type geodesics. These oscillate between two angular turning points either in the northern or southern hemisphere without ever crossing the equatorial plane. Vortical geodesics, corresponding to a limited region within the critical curve on the image plane, will not be considered further in our analysis~\cite{Gralla:2019drh}.

For a light ray with $\eta>0$, combining $\tau \equiv G_\theta$ with Eq.~(\ref{eq:Gtheta}), the final polar angle $\theta_f$ can be elegantly expressed in terms of the Mino time $\tau$~\cite{Gralla:2019ceu}
\begin{equation}
    \frac{\cos\theta_f}{\sqrt{u_+}}=-\nu_{\theta}^s\operatorname{sn}\left(\sqrt{-u_-a^2}\big(\tau+\nu_{\theta}^s\G_{\theta}^s\big)\;\middle|\;\frac{u_+}{u_-}\right),
    \label{eq:theta_f}
\end{equation}
where $\operatorname{sn}$ denotes the Jacobi elliptic sine function. Using $m$ to represent the number of angular turning points the light ray encounters, the $G$-integrals $G_j\, (j=\theta,\phi,t)$ can be articulated in a universal manner~\cite{Gralla:2019ceu}
 \begin{equation}
 G_j=m\left(\G_j^+-\G_j^-\right)+\nu_{\theta}^s\left[(-1)^m\G_j^f-\G_j^s\right],
 \label{eq:angular int}
 \end{equation}
where the superscripts ${+,-,f,s}$ correspond to the evaluation of $\G_j$ at $\theta=\theta_+$, $\theta_-$, $\theta_f$, and $\theta_s$, respectively. Ref.~\cite{Gralla:2019ceu} provides the analytic expressions for $\G_j$.

\paragraph{Radial motion}
There exists a family of bound orbits where geodesics maintain a constant radius $\tilde{r}$. This radius corresponds to double roots of the radial potential $\mathcal{R}(r)$, satisfying
$\mathcal{R}(\tilde{r})=\mathcal{R}^{\prime}(\tilde{r})=0$.  For Kerr spacetime, the possible values of $\tilde{r}$ fall within the range $\tilde{r}\in[\tilde{r}_-,\ \tilde{r}_+]$, where
    \begin{equation}
        \tilde{r}_{\pm} \equiv 2M\bigg[1+\cos\bigg(\frac{2}{3}\arccos\left(\pm\frac{a}{M}\right)\bigg)\bigg],
        \label{eq:bound radius range}
    \end{equation}
and bound orbits are characterized by conserved quantities as functions of $\tilde{r}$
    \begin{equation}
        \tilde{\lambda}=a+\frac{\tilde{r}}{a}\left[\tilde{r}-\frac{2\Delta(\tilde{r})}{\tilde{r}-M}\right], \quad
        \tilde{\eta}=\frac{\tilde{r}^3}{a^2}\left[\frac{4M\Delta(\tilde{r})}{(\tilde{r}-M)^2}-\tilde{r}\right]. \label{eq:lamda eta c}
    \end{equation}
Photons with such conserved quantities are said to be ``critical'', and approach the bound orbit at $\tilde{r}$ in the asymptotic future or past~\cite{Gralla:2019drh}.

Photon bound orbits are inherently unstable: any small radial deviation will grow exponentially, leading the photon to either fall into the BH or escape to infinity. 
Specifically, consider a critical orbit with conserved quantities $(\tilde{\lambda}(\tilde{r}), \tilde{\eta}(\tilde{r}))$, and a deviation from its corresponding bound orbit denoted by $r = \tilde{r} + \delta r_0$, where $\left|\delta r_0\right| \ll M$. After $n$ half-librations in the $\theta$ direction, the photon arrives at a radius $\tilde{r} + \delta r_n$, where the radial deviation grows exponentially as
\begin{equation}
    \frac{\delta r_n}{\delta r_0}\approx \ee^{n\,\gamma},
\end{equation}
where $\gamma$ is the Lyapunov exponent characterizing the instability of bound orbits, as defined in Refs.~\cite{Cardoso:2008bp,Gralla:2019drh}. Conserved quantities of critical photons correspond to a curve known as the ``critical curve'' on the $\lambda$-$\eta$ plane, which delineates two distinct regions in the positive-$\eta$ portion of the $\lambda$-$\eta$ plane, each corresponding to a different type of radial motion.

The classification of radial motion leverages the roots of the radial potential $\mathcal{R}(r)$, denoted as ${r_1, r_2, r_3, r_4}$, arranged such that
$\operatorname{Re}r_1\leq\operatorname{Re}r_2\leq\operatorname{Re}r_3\leq\operatorname{Re}r_4$. Detailed analytic expressions for these roots are available in Ref.~\cite{Gralla:2019ceu}. The behavior of $r_4$ is crucial in determining the nature of radial motion. When $(\lambda, \eta)$ falls inside the critical curve, $r_4$ is either complex or real but less than the outer horizon at $r_h \equiv M + \sqrt{M^2 - a^2}$. In this scenario, the radial potential $\mathcal{R}(r)$ does not admit zeros outside the horizon, indicating no radial turning points are present, and only photons with $\nu_r^s = +1$ can escape. We denote the number of radial turning points encountered by $w$, which is zero ($w = 0$) in this scenario.

Conversely, for $(\lambda, \eta)$ outside the critical curve, $\mathcal{R}(r)$ exhibits four real roots, with only two outside the horizon satisfying $r_4 > r_3 > r_h$. According to Eq.~(\ref{eq:pr}), radial motion is permissible where $\mathcal{R}(r) \geq 0$, specifically in the range $r_h < r \leq r_3$ or $r_4 \leq r < \infty$. If the source radius $r_s$ is within the first interval, escape is impossible for the photon. Consequently, only scenarios where $r_s > r_4$ are considered relevant, accommodating both $\nu_r^s = +1$ (direct escape, $w = 0$) and $\nu_r^s = -1$ (bounce-back at $r_4$, $w = 1$). The various types of valid radial motions described are succinctly summarized in Table~\ref{tab:radial types}.

\begin{table}[!htbp]
    \renewcommand\arraystretch{1.3} %% row spacing
    \setlength\tabcolsep{4pt} %% column spacing
    \begin{tabular}{cccc}\hline\hline
       Case & $r_4$ & Condition & $w$ \\ \hline
        inside CC & $r_4\notin\mathbb{R}$ or $r_4<r_h$ & $\nu_r^s=+1$  & $0$ \\ 
       \multirow{2}{*}{outside CC} &  \multirow{2}{*}{$r_4>r_h$} & $r_s>r_4,\ \nu_r^s=+1$ & $0$ \\ 
        & & $r_s>r_4,\ \nu_r^s=-1$ & $1$\\ \hline
    \end{tabular}
\caption{Classification of radial motion capable of escaping the BH, delineated by the number $w$ of radial turning points encountered by the photon. CC denotes the critical curve.}
    \label{tab:radial types}
\end{table}

For rays that escape the BH and eventually reach a radius $r_f \gg M$, the $I$-integrals $I_i$ (where $i=r,\phi,t$) along their trajectory can be formulated as
    \begin{equation}
        I_i=\I_i^f-\I_i^s+2w\left(\I_i^s-\I_i^4\right),
    \label{eq:radial int}
    \end{equation}
with the superscripts $f$, $s$, and $4$ indicating the evaluation of $\I_i$ at $r=r_f$, $r_s$, and $r_4$, respectively. Analytic formulas for the antiderivative $\I_i(r)$ (for $i=r,\phi,t$) are detailed in Ref.~\cite{Gralla:2019ceu}.

%%%%%%%%%%%%%%%%%%%%%%%%%%%%%%%%%%%%%%%%%
\subsection{Image Plane}\label{subsec:OP}
%%%%%%%%%%%%%%%%%%%%%%%%%%%%%%%%%%%%%%%%%
For the observational analysis, we consider light rays that have escaped the vicinity of the BH and reached a distant observer located at position $(r_o, \theta_o, \phi_o)$, where $r_o \gg M$. At this remote distance, the trajectory of the light ray is nearly straight as it approaches the observer. Tracing it back as a straight line, we can quantify its perpendicular deviation from the BH using two-dimensional impact parameters $(\alpha, \beta)$~\cite{Bardeen:1973tla,Gralla:2017ufe,Gralla:2019drh}
\begin{equation}
    \begin{aligned}
    \alpha&=-\frac{\lambda}{\sin\theta_o},\\
    \beta&=\nu_{\theta}^o\sqrt{\Theta(\theta_o)}=\nu_{\theta}^o\sqrt{\eta+a^2\cos^2\theta_o-\lambda^2\cot^2\theta_o},
    \label{eq:impact parameters}
    \end{aligned}
\end{equation}
where $\nu_{\theta}^o$ denotes the sign of $p^{\theta}$ upon the ray's arrival at the observer. These parameters define a two-dimensional orthogonal coordinate system on the image plane, which will be illustrated in the following section in the top left panel of Fig.~\ref{fig:images}. Consequently, the critical curve on the $\lambda$-$\eta$ plane translates into a corresponding critical curve on the $\alpha$-$\beta$ plane, with the upper and lower halves representing $\nu_{\theta}^o = \pm 1$, respectively. The projection of the BH's spin aligns with the $\beta$-axis, and the observer's direct line of sight to the BH is marked by $\alpha = \beta = 0$. By dividing $\alpha$ and $\beta$ by the distance from the BH to the observer, we obtain two direction cosines on the observer's sky, indicating the apparent angular position of the image.

In practice, simulating an image of a BH surrounded by an accretion flow often involves backward ray tracing. This method launches geodesics from a grid on the observer's image plane, parametrized by $(\alpha, \beta)$, back toward the BH~\cite{Bardeen:1973tla,Luminet:1979nyg}. The geodesics' endpoints either fall into the BH or extend to a large Mino time for photons situated outside or on the critical curve. The next step involves calculating the emissivity and absorption coefficients along each geodesic. These values are then integrated back to the observer, factoring in both gravitational and Doppler redshift effects, in a procedure termed covariant radiative transfer~\cite{Gammie_2012, Dexter:2016cdk}.

%%%%%%%%%%%%%%%%%%%%%%%%%%%%%%%%%%%%%%%%
\section{Forward Ray Tracing and Light Echoes}\label{sec:FRLE}
%%%%%%%%%%%%%%%%%%%%%%%%%%%%%%%%%%%%%%%%
The backward ray tracing method, while effective in many applications, is inefficient for simulating higher-order images from pointlike sources such as hotspots. This inefficiency arises from the challenge of aligning light rays from a finite observer grid precisely with the emission point. Moreover, the photon ring's inherent instability further complicates this process, as the probability of correctly capturing these light rays decreases exponentially with increasing image order $N$. To address these challenges and facilitate the simulation of hotspot images, we employ forward ray tracing~\cite{Cunningham1973,Vazquez:2003zm}. Utilizing the analytic geodesic solutions discussed in Sec.~\ref{subsec:Trajectory} and an optimized parametrization scheme for conserved quantities, our forward ray tracing method effectively captures multiple images from a point source.

%%%%%%%%%%%%%%%%%%%%
\subsection{Forward Ray Tracing}
\label{subsec:frt}
%%%%%%%%%%%%%%%%%%%%
This subsection elaborates on the methodology behind our forward ray tracing approach. This approach, initiating at the source point, computes light rays that ultimately converge at the observer, yielding observable images on the plane described in Sec.~\ref{subsec:OP}. Note that our discussion is confined to scenarios where $\eta > 0$, thus omitting vortical motion.

\subsubsection{Boundary Conditions to Fix Light Rays}
\label{subsubsec:direction matching}

The core principle of forward ray tracing is to determine the paths of light rays connecting a given emission point, specified by coordinates $(r_s, \theta_s, \phi_s)$, with a distant observer at $(r_o, \theta_o, \phi_o)$. These light rays are characterized by their conserved quantities, $\lambda$ and $\eta$, along with the directional signs of their momenta, $\nu_r^s$ and $\nu_{\theta}^s$, allowing them to complete multiple orbits around the BH before reaching the observer.

Before identifying feasible light rays by solving for permissible combinations of $(\lambda, \eta, \nu_r^s, \nu_{\theta}^s)$, we first delineate the allowed parameter space using two necessary conditions. The initial condition mandates that the polar angle of the source, $\theta_s$, falls within the angular motion range $[\theta_-, \theta_+]$. Pairs of $(\lambda, \eta)$ leading to a range of angular motion that does not include $\theta_s$ are excluded. The second condition assesses the photon's ability to escape the BH, achieved by calculating $r_4$ and verifying the escape criterion outlined in the third column of Table~\ref{tab:radial types}.

Within the refined parameter space, each $(\lambda, \eta, \nu_r^s, \nu_{\theta}^s)$ combination corresponds to a light ray directed toward a distant observer. Next, we consider the light ray's final direction. We begin by setting the final radial position to $r_f = r_o$, which allows us to determine the Mino time $\tau \equiv I_r$ through Eq.~(\ref{eq:radial int}). Subsequently, the final polar angle $\theta_f$ is obtained from Eq.~(\ref{eq:theta_f}). This enables us to calculate the number of angular turnings $m$ encountered during propagation
\begin{equation}
m = 1 + \left\lfloor \frac{I_r - \G_{\theta}^{+} + \nu_{\theta}^s\G_{\theta}^s}{\G_{\theta}^+ - \G_{\theta}^-} \right\rfloor,
\label{eq:m}
\end{equation}
where $\lfloor x \rfloor$ denotes the largest integer less than or equal to $x$. This expression can be derived as follows: the elapsed Mino time before encountering the first turning point is $(\G_{\theta}^+-\G_{\theta}^s)$ for $\nu_{\theta}^s=+1$ and $(\G_{\theta}^s-\G_{\theta}^-)$ for $\nu_{\theta}^s=-1$, which can be combined into
$(\G_{\theta}^{+} - \nu_{\theta}^s \G_{\theta}^s)$ using the relation $\G_{\theta}^- = -\G_{\theta}^+$~\cite{Gralla:2019ceu}. The rest of the journey corresponds to Mino time $(I_r - \G_{\theta}^{+} + \nu_{\theta}^s \G_{\theta}^s)$, which gives the additional turning number by dividing by $(\G_{\theta}^+ - \G_{\theta}^-)$ and taking the integer part. Incorporating $\theta_f$ and $m$ into Eq.~(\ref{eq:angular int}) yields $G_\phi$, and $I_{\phi}$ is derived from Eq.~(\ref{eq:radial int}). These combined enable the determination of the final azimuthal angle $\phi_f$ through Eq.~(\ref{eq:inteq2}).

The concluding requirement for a light ray is to satisfy the imaging conditions~\cite{Cunningham1973}
\begin{equation}
\begin{aligned}
    \theta_f &= \theta_o, \\ 
    \phi_f &= \phi_o + 2k\pi\ (k\in\mathbb{Z}).
\end{aligned}
\label{eq:imaging condition}
\end{equation}
Consequently, identifying images of a point source is effectively reduced to solving a two-dimensional root-finding problem for $(\lambda, \eta)$ individually for four types of momentum sign combinations $(\nu_r^s, \nu_{\theta}^s) = (\pm 1, \pm 1)$.

\begin{figure}[!htb]
    \centering
    \includegraphics[width=0.48\textwidth]{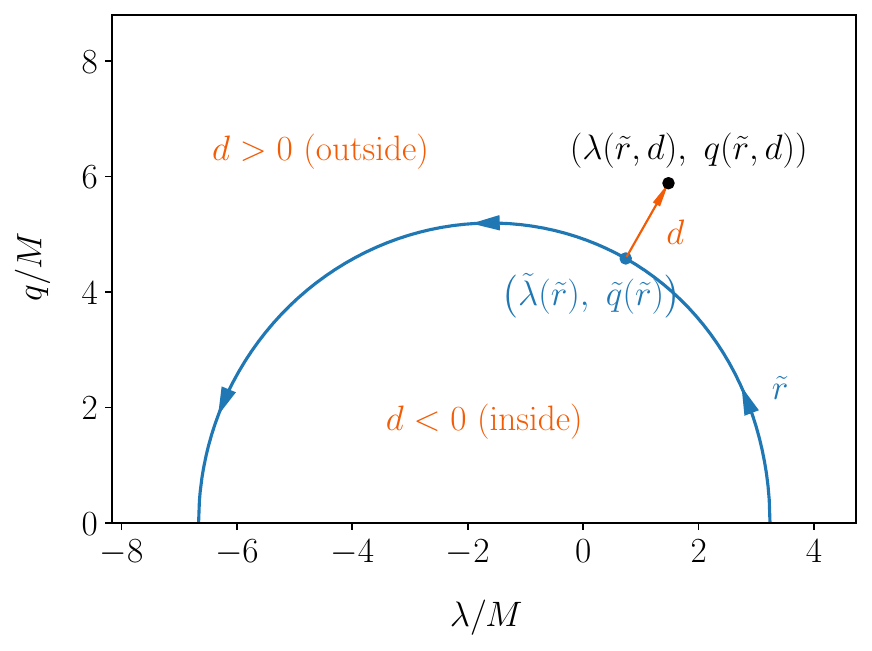}
    \caption{Parametrization of conserved quantities of geodesics used in root-finding. The example shown is for a BH with $a/M = 0.8$. The blue curve represents the critical curve corresponding to bound orbits, with arrows indicating the direction of increasing $\tilde{r}$.}
    \label{fig:rc_d parameterization}
\end{figure}

In practice, to efficiently identify the roots with high precision, we transform the variables $(\lambda, \eta)$ into a new pair of parameters. We begin by introducing $q \equiv \sqrt{\eta}$, ensuring $\lambda$ and $q$ share the same dimensionality. The value of $q$ along the critical curve is denoted as $\tilde{q} \equiv \sqrt{\tilde{\eta}}$. We then assess the deviation of $(\lambda, q)$ from their critical curve counterparts $(\tilde{\lambda}, \tilde{q})$ via the parametrization
\begin{equation}
    \begin{pmatrix}
        \lambda \\
        q
    \end{pmatrix} \equiv \begin{pmatrix}
        \tilde{\lambda} \\ 
        \tilde{q}
    \end{pmatrix} + \vec{n}_{\lambda q} d,
\end{equation}
where
\begin{equation}
\begin{aligned}
        \vec{n}_{\lambda q} &\equiv \frac{(\p_{\tilde{r}}\tilde{q},\ -\p_{\tilde{r}}\tilde{\lambda})^{\mathrm{T}}}{\sqrt{(\p_{\tilde{r}}\tilde{\lambda})^2+(\p_{\tilde{r}}\tilde{q})^2}} \\ &= \frac{ \left[\tilde{r}^2(3M-\tilde{r}),\  a\tilde{q}(\tilde{r}-M)\right]^{\mathrm{T}}} {\sqrt{\tilde{r}^4(3M-\tilde{r})^2+a^2\tilde{q}^2(\tilde{r}-M)^2}}
\end{aligned}
\end{equation}
represents the outward-pointing unit normal vector to the critical curve $(\tilde{\lambda}, \tilde{q})$ on the $\lambda$-$q$ plane. The parameter $d$ indicates the distance of $(\lambda, q)$ from the critical curve, with regions outside and inside the curve distinguished by $d > 0$ and $d < 0$, respectively. This approach allows each point $(\lambda, q)$ to be uniquely mapped to a distance $d$ and the radius $\tilde{r}$ of the corresponding critical curve, as illustrated in Fig.~\ref{fig:rc_d parameterization}. Given that higher-order images asymptotically approach the critical curve, we opt for $\log_{10}|d/M|$ as the variable to accommodate the asymptotic behavior of radial motion. This parametrization scheme effectively alleviates the numerical challenges arising from the exponentially increased sensitivity of $(\theta_f, \phi_f)$ on conserved quantities as higher-order images approach the critical curve. The root-finding process is then transformed from the original $\lambda$-$\eta$ plane to the $\tilde{r}$-$\log_{10}|d/M|$ plane, with a total of eight possible combinations of sign variables $(\nu_r^s, \nu_{\theta}^s, \operatorname{sgn}(d)) = (\pm1, \pm1, \pm1)$, which should be solved one by one. For each combination of sign variables, the contours for $\theta_f = \theta_o$ and $\phi_f = \phi_o + 2k\pi\ (k\in\mathbb{Z})$ are drawn, with their intersections denoting the images. Note that for a Schwarzschild BH, since $\tilde{r}$ is always $3M$, finding a root in the one-dimensional space of $\log_{10}|d/M|$ suffices. Equivalently, this can be done by restricting to the equatorial plane, setting $\eta = 0$, and solving for the remaining conserved quantity $\lambda$. See Sec.~\ref{subsec:Sch distortion} for details.

Our forward ray tracing method is similar to that used in Ref.~\cite{Vazquez:2003zm}, which employs analytic geodesic solutions for computational efficiency and accuracy. However, that study only considers a point source located infinitely far from the BH ($r_s \to \infty$), leading to lensed images (referred to as relativistic images therein) originating solely from the region outside the critical curve in the $\lambda$-$\eta$ plane. In this work, we generalize this methodology to arbitrary source positions, utilizing comprehensive analytic geodesic solutions that account for parameter spaces both inside and outside the critical curve. For instance, in Fig.~\ref{fig:images for various spins}, images appear inside the critical curve for $r_s = 1.7\,M$ and outside for $r_s = 10\,M$.

\subsubsection{Light Echoes}
\label{subsubsec:light echoes}
Generally, starting from a source at $x^{\mu}_s$, there are infinitely many light rays satisfying Eq.~(\ref{eq:imaging condition}), forming a sequence of images on the observer's image plane. Before reaching the observer, each of these light rays revolves around the BH for a different number of turns. One way to label these light echoes is by introducing the number of half-orbits propagated in the $\theta$ direction, denoted by $n$
\begin{equation}
    n\equiv\frac{G_{\theta}}{G_{\theta}^{\text{half}}},\qquad G_{\theta}^{\text{half}}\equiv\int_{\theta_-}^{\theta_+}\frac{\dd\theta}{\sqrt{\Theta(\theta)}},
    \label{eq:n}
\end{equation}
which differs from the definition in Ref.~\cite{Gralla:2019drh} by a factor of $2$. The noninteger number $n$ quantifies the length of the journey in the $\theta$ direction experienced by the photon. Every time $n$ increases by two, the photon experiences one more full oscillation in $\theta$. This definition can be connected to the integer definition of the number of half-orbits, $N$, through
\begin{equation}
    N=\left\lfloor n \right\rfloor,
    \label{eq:N}
\end{equation}
which suitably defines the level of each image.
For the \textit{direct image}, which is defined as the one whose trajectory undergoes the least bending from gravity, we have {$n < 1$} and its level is $N = 0$. The following higher-order images have $N=1,\,2,\,3\ldots$, and we call them \textit{lensed images}. Higher-order images with larger $N$ asymptotically approach the critical curve on the $\alpha$-$\beta$ plane. In the following, we will mainly use the level number $N$ to label each image. If multiple images have the same level $N$, we distinguish them by sorting in order of increasing $n$ and assigning an adjunctive letter to each in alphabetical order $(a,\,b,\,c\ldots)$. For example, if there are three images at level $N = 7$, we use $7a$, $7b$, and $7c$ to denote them, with their number of {half-}orbits $n$ satisfying $n(7a) < n(7b) < n(7c)$.

%%%%%%%%%%%%%%%%%%%%%%%%%%%%%
\subsubsection{An Example}
\label{subsubsec:example}
%%%%%%%%%%%%%%%%%%%%%%%%%%%%%%
To illustrate the application of our forward ray tracing method, we consider an example: a hotspot located at $(r_s=10\,M, \theta_s=\ang{90}, \phi_s=-\ang{45})$ outside a BH with spin $a=0.8\,M$, and an observer positioned at $(r_o=1000\,M, \theta_o=\ang{17}, \phi_o=0)$. Note that $r_o = 1000\,M$ is far enough that for larger distances, the solutions for $(\lambda, \eta)$ remain nearly unchanged.

\begin{figure*}[p]
    \centering
    \includegraphics[width=1\textwidth]{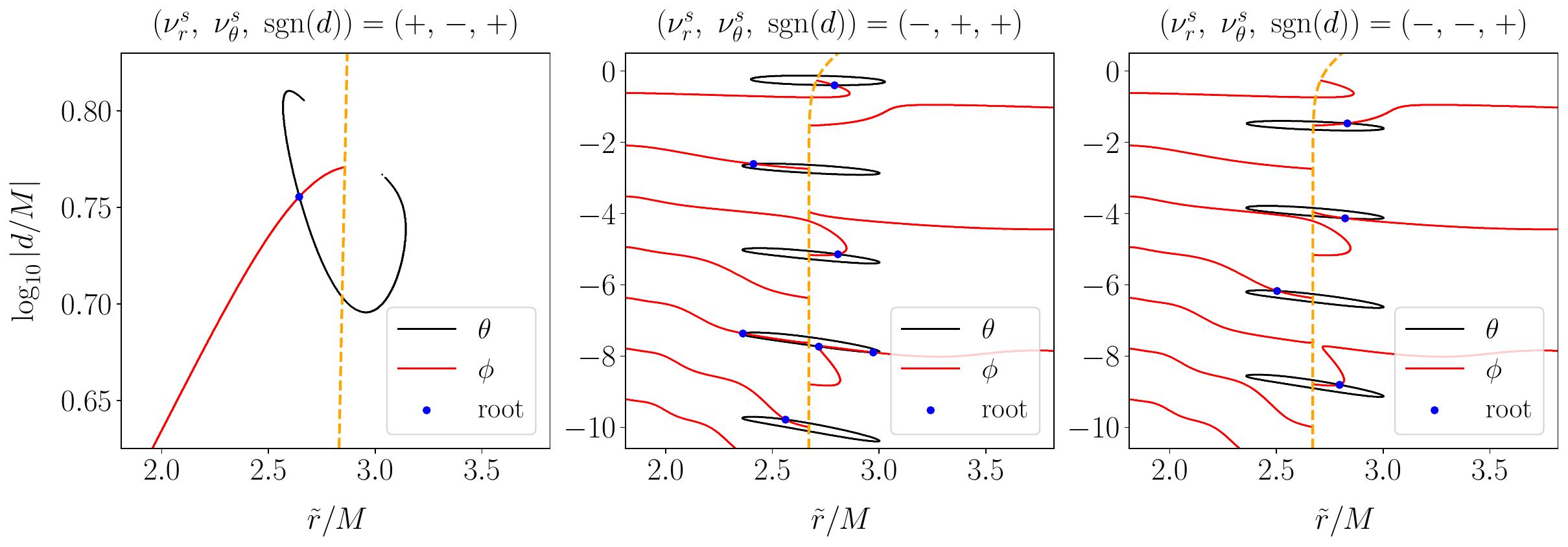}
    \caption{Contour plots illustrating the root-finding process in forward ray tracing for the example discussed in Sec.~\ref{subsubsec:example}: a pointlike source located at $(r_s=10\,M$, $\theta_s=\ang{90}$, $\phi_s=-\ang{45})$ outside a BH with spin $a=0.8\,M$. The observer is positioned at
    $(r_o=1000\,M$, $\theta_o=\ang{17}$, $\phi_o=0)$. The yellow dashed lines at $\lambda = 0$ divide the plane into regions with opposite azimuthal directions: $\phi_f - \phi_s > 0$ (left) and $\phi_f - \phi_s < 0$ (right). Black and red lines represent the solutions $\theta_f = \theta_o$ and $\phi_f = \phi_o + 2k\pi$ ($k\in\mathbb{Z}$), respectively, and the blue points at their intersections correspond to the light ray solutions. Three sets of solutions are shown: direct emission (left), and lensed photons with contrasting signs of $\nu_{\theta}^s$ (middle and right).}
    \label{fig:contour plots}
\end{figure*}

We first present in Fig.~\ref{fig:contour plots} the contour plots demonstrating the root-finding process. Yellow dashed lines mark $\lambda = 0$, separating the plane into two regions of opposite azimuthal directions: $\phi_f - \phi_s > 0$ (left) and $\phi_f - \phi_s < 0$ (right). Black and red lines correspond to the solutions $\theta_f = \theta_o$ and $\phi_f = \phi_o + 2k\pi\ (k\in\mathbb{Z})$, respectively, and blue points are positioned at their intersections, corresponding to the solutions found in the forward ray tracing. Our analysis identifies three sets of solutions that satisfy the imaging condition Eq.~(\ref{eq:imaging condition}), organized by values of $(\nu_r^s, \nu_{\theta}^s, \operatorname{sgn}(d))$ and listed in Table~\ref{tab:roots}. The image positions $(\alpha, \beta)$, along with the arrival time $t_f$ assuming emission time $t_s = 0$, are also detailed in Table~\ref{tab:roots}.

\begin{figure*}[p]
    \centering
    \begin{minipage}{0.45\textwidth}
        \includegraphics[width=0.95\textwidth]{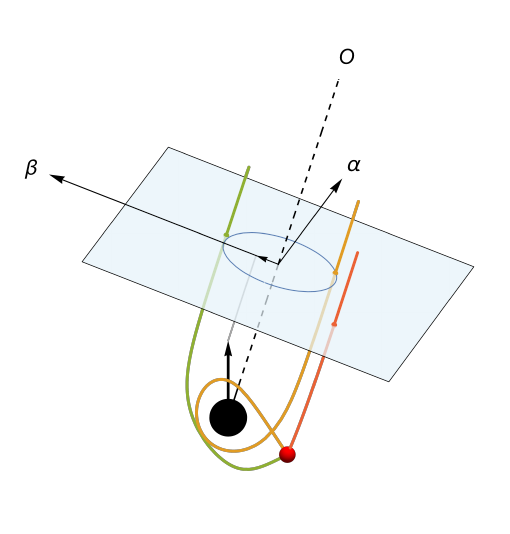}
    \end{minipage}
    \quad
    \begin{minipage}{0.45\textwidth}
        \includegraphics[width=0.95\textwidth]{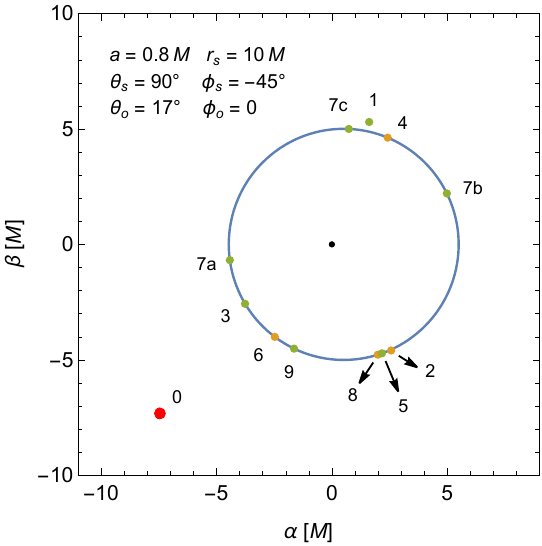}
    \end{minipage}
    \includegraphics[width=0.8\textwidth]{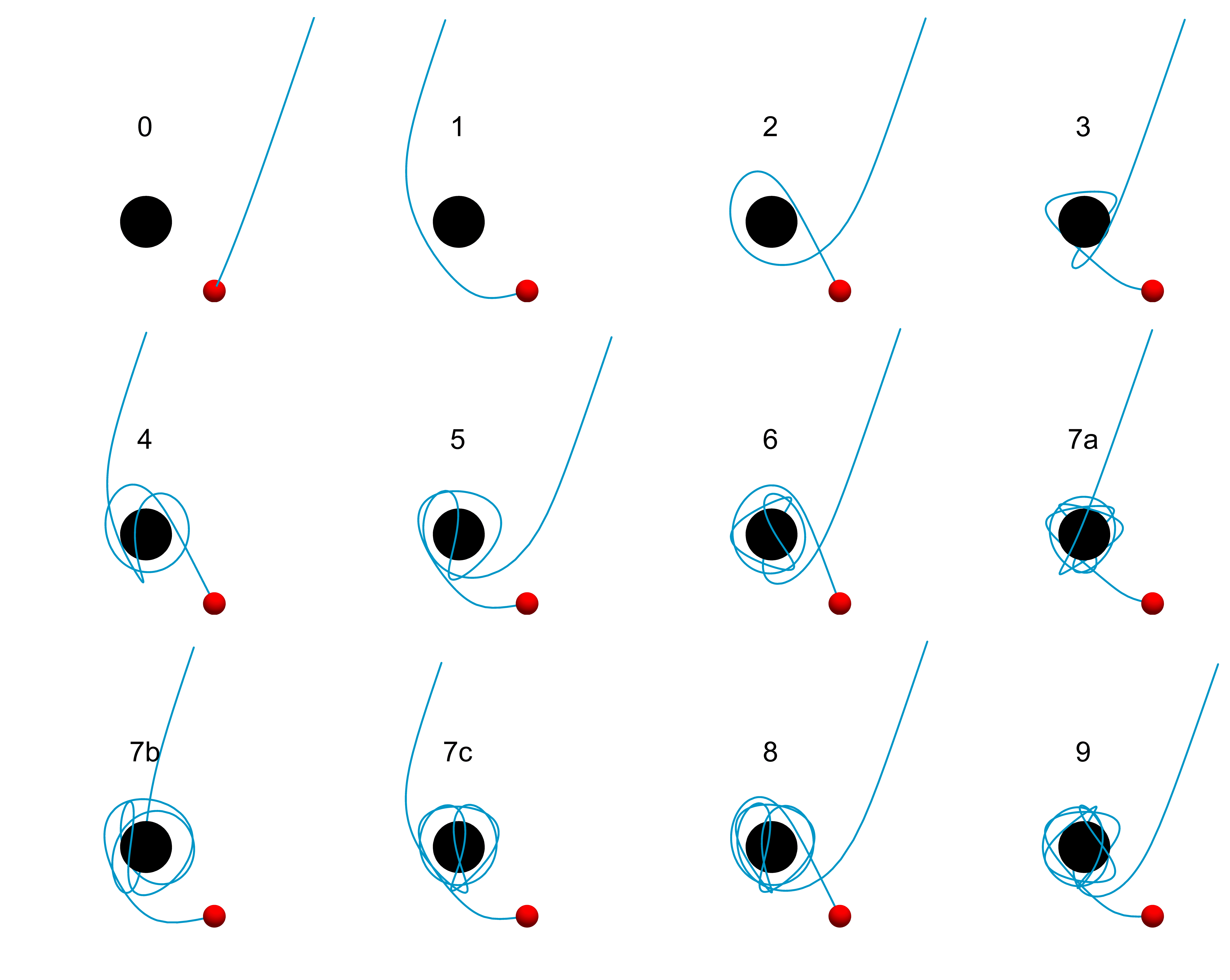}
    \caption{Geodesics and images of the solutions found in forward ray tracing for the example discussed in Sec.~\ref{subsubsec:example}.
    \textit{Top left}: Illustration of the image plane $\alpha-\beta$ showing the solutions of the lowest three levels. The dashed black line perpendicular to the plane represents the line of sight from the observer to the BH. The short black arrow on the BH denotes its spin direction, with its projection aligning with the $\beta$-axis on the plane. Light rays connect the pointlike source (red sphere) with the faraway observer ($O$).
    \textit{Top right}: Images from the pointlike source on the image plane centered at the line-of-sight direction of the BH (black point). The direct image is shown in red, while the lensed images are in green for $\nu_{\theta}^s = +1$ and orange for $\nu_{\theta}^s = -1$. The critical curve is plotted in blue.
    \textit{Bottom}: Trajectories of the light rays that form the images for $N$ from $0$ to $9$.}
    \label{fig:images}
\end{figure*}

In Fig.~\ref{fig:images}, we present the corresponding geodesics and images. The top left panel illustrates the image plane introduced in Sec.~\ref{subsec:OP}, with various images depicted in the top right panel. The bottom panel shows the trajectories corresponding to each order of images.

\begin{table*}[p]
\centering
\renewcommand\arraystretch{1.3} %% row spacing
\setlength\tabcolsep{5pt} %% column spacing
\begin{tabular}{ccc S[table-format=1.5] S[table-format=2.5] c S[table-format=2.2] S[table-format=2.2] S[table-format=4.2] S[table-format=1.3] c}
\hline\hline
\multirow{2}{*}{Type}  & {\hspace{1.8em}}  & \multicolumn{3}{c}{Root} & {\hspace{1.8em}} & \multicolumn{5}{c}{Image} \\ 
                       &    & $\left(\nu_r^s,\ \nu_{\theta}^s,\ \operatorname{sgn}(d)\right)$ & {$\tilde{r}/M$} & {$\log_{10}\left|d/M\right|$} &    & {$\alpha/M$} & {$\beta/M$} & {$t_f/M$} & {$n$} & $N$ \\ \hline
direct image:          &    &                 &         &          &    &       &       &         &       &   \\ 
                       &    &  $(+,\ -,\ +)$  & 2.64422 &  0.75554 &    & -7.45 & -7.32 & 1007.81 & 0.433 & $0$ \\
lensed images:         &    &                 &         &          &    &       &       &         &       &   \\   
                       &    &  $(-,\ +,\ +)$  & 2.79133 & -0.40197 &    &  1.62 &  5.30 & 1037.38 & 1.590 & $1$ \\  
                       &    &                 & 2.41127 & -2.61071 &    & -3.76 & -2.58 & 1066.95 & 3.446 & $3$ \\  
                       &    &                 & 2.80707 & -5.14474 &    &  2.17 & -4.72 & 1097.41 & 5.414 & $5$ \\  
                       &    &                 & 2.36197 & -7.36752 &    & -4.42 & -0.68 & 1130.81 & 7.485 & $7a$ \\  
                       &    &                 & 2.97236 & -7.89958 &    &  4.98 &  2.21 & 1131.15 & 7.539 & $7b$ \\  
                       &    &                 & 2.71788 & -7.73576 &    &  0.74 &  5.00 & 1131.26 & 7.593 & $7c$ \\  
                       &    &                 & 2.56144 & -9.78463 &    & -1.64 & -4.52 & 1159.67 & 9.411 & $9$ \\ 
                       &    &  $(-,\ -,\ +)$  & 2.83263 & -1.47102 &    &  2.57 & -4.60 & 1050.67 & 2.417 & $2$ \\  
                       &    &                 & 2.82223 & -4.13481 &    &  2.42 &  4.62 & 1084.52 & 4.584 & $4$ \\  
                       &    &                 & 2.50343 & -6.17371 &    & -2.47 & -4.01 & 1113.21 & 6.420 & $6$ \\  
                       &    &                 & 2.79631 & -8.80814 &    &  1.99 & -4.78 & 1144.13 & 8.413 & $8$ \\  \hline
\end{tabular}
\caption{List of roots found with $N \leq 9$ in the forward tracing for the example discussed in Sec.~\ref{subsubsec:example}. The roots are classified into three categories based on different combinations of $(\nu_r^s,\,\nu_{\theta}^s,\,\operatorname{sgn}(d))$. The table displays their image positions $(\alpha, \beta)$, arrival times $t_f$ (assuming the emission time $t_s = 0$), and the number of {half-}orbits propagated in the $\theta$ direction $n$. Each root is distinguished by its level number $N$.}
\label{tab:roots}
\end{table*}

\begin{figure*}[p]
    \centering
    \includegraphics[width=\textwidth]{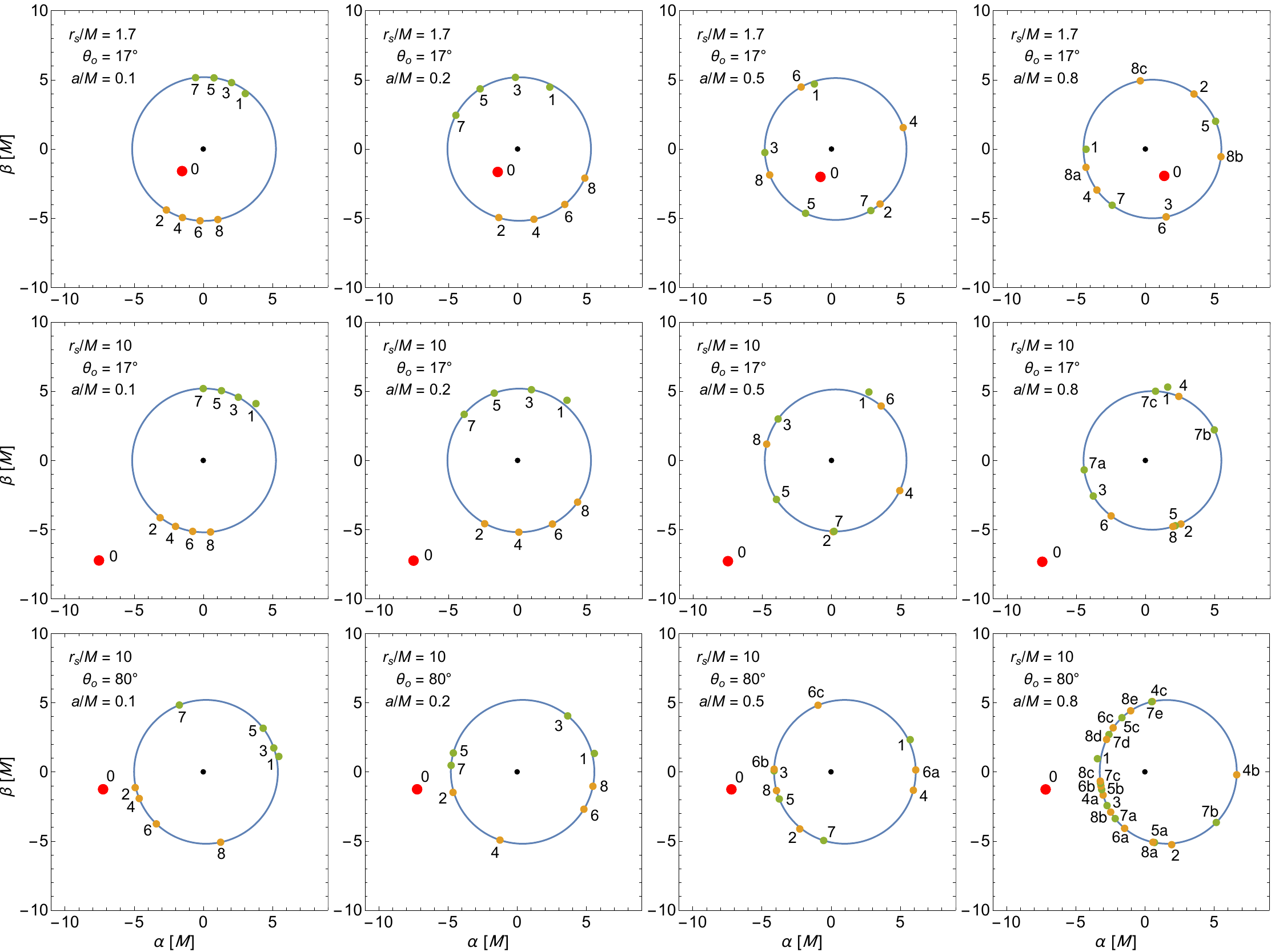}
    \caption{
    Forward ray tracing images from a point source at various source locations $(r_s,\,\theta_s,\,\phi_s)$, BH spin $a$, and observer inclination angles $\theta_o$. Fixed parameters across all cases are $\theta_s = \ang{90}$, $\phi_s = -\ang{45}$, and $\phi_o = 0$. The first row shows a source at $r_s = 1.7\,M$, while the second and third rows depict sources at $r_s = 10\,M$, with inclination angles of $\theta_o = \ang{17}$ and $\theta_o = \ang{80}$, respectively. The spin parameter $a/M$ varies from $0.1, 0.2, 0.5$, to $0.8$ across the columns from left to right. The labeling and coloring conventions for the images follow those used in the top right panel of Fig.~\ref{fig:images}. To enhance clarity, particularly in densely populated panels, labels for odd-level images (green dots) are positioned inside the critical curve, whereas labels for even-level lensed images (orange dots) are placed outside. Where image dots overlap, higher-level images are superimposed on lower-level ones.}
    \label{fig:images for various spins}
\end{figure*}

Since the source is in the foreground of the BH relative to the observer, the direct image arises from a light ray experiencing minimal gravitational bending, containing momentum signs $\nu_r^s = +1$ and $\nu_\theta^s = -1$, straightforwardly reflecting the spatial configuration between the source and the observer. Lensed images, on the other hand, are categorized into two types. For both types, $\nu_r^s = -1$ is observed because the source is located outside the region containing bound orbits, i.e., $r_s$ is greater than $\tilde{r}_+ \approx 3.82\,M$ as defined in Eq.~(\ref{eq:bound radius range}). These two categories are distinguished by their contrasting values of $\nu_\theta^s$. The sequential decrease in the $\log_{10}|d/M|$ values of the roots demonstrates that higher-order images are closer to the critical curve.

The fact that lensed images ($N \geq 1$) can be categorized into two sets is well expected and can be explained as follows. Lensed images are all formed by light rays carrying conserved quantities close to the critical curve. Their orbits, when propagating near the bound photon orbit, are nearly spherical. In the Schwarzschild case, nearly critical photons enter these quasibound orbits in two distinct directions: clockwise or counterclockwise. Therefore, we expect the initial photon directions of lensed images to naturally separate into two clusters. We find this phenomenon still widely holds in Kerr spacetime, despite the effect of spin that complicates the motion of photons. In the example here, the two initial photon clusters are distinguished by contrasting $\nu_\theta^s$, with odd and even level numbers $N$ in each category. Note that at the $N = 7$ level, there are three different images, and therefore we label them as $7a$, $7b$, and $7c$. In fact, the appearance of multiple images with the same $N$ is common in Kerr spacetime, which will discussed further in Sec.~\ref{subsec:image degeneracy}.

%%%%%%%%%%%%%%%%%%%%%%%%%%%%%%%%%%%%%%%%
\subsection{Time Delay and Position Angle Shift of Light Echoes}
\label{subsec:delay and shift}
%%%%%%%%%%%%%%%%%%%%%%%%%%%%%%%%%%%%%%%%
We have demonstrated an example of forward ray tracing in Fig.~\ref{fig:contour plots}, Table~\ref{tab:roots} and Fig.~\ref{fig:images}. Further examples of image planes for various BH parameters and source location are shown in Fig.~\ref{fig:images for various spins}. From left to right, we choose the spin $a/M$ to be $0.1, 0.2, 0.5$ and $0.8$, respectively. The first row considers a source close to the BH with $r_s = 1.7\,M$, resulting in $\nu_r^s = +1$ for all lensed images as $r_s < \tilde{r}_{-}$. The last two rows consider the source located at $r_s = 10\,M$, with the second row for nearly face-on observation ($\theta_o = \ang{17}$) and the third for nearly edge-on observation ($\theta_o = \ang{80}$).

For observational purposes, the most relevant quantities are the position and arrival time of light echoes, which encode information about the spacetime and the source location. One way to measure these observables is through intensity fluctuation correlation \cite{Broderick:2005my,Fukumura:2007xr,Moriyama:2015zfa,Saida:2016kpk,Gralla:2017ufe,Moriyama:2019mhz,Tiede:2020jgo,Hadar:2020fda,Hadar:2023kau}, where intrinsic time variations of the emission source can be detected at different image levels, delayed by certain amounts of time.

Comparing two images with sequential levels, their propagation distance does not necessarily differ by a half-orbit of $\theta$, as shown by the $n$ values in Table~\ref{tab:roots}. The difference in $n$ between the two is sensitive to the source location. Consequently, the time delay and position angle shift between the direct ($N = 0$) and the first-lensed ($N = 1$) images depend on both the BH and the source.

On the other hand, images with levels differing by $2$ are usually separated by one orbit of $\theta$, since they share a similar initial direction characterized by $\nu_\theta^s$, as seen in Table~\ref{tab:roots}. Consequently, the time delay between the $N$-th and the $(N + 2)$-th images is mostly determined by the universal properties of the BH
\begin{equation}
t_f^{N+2} - t_f^N \approx 2\tau,
\end{equation}
where $\tau \approx 15\,M$ is the critical parameter characterizing the temporal period of a half-libration in the $\theta$ direction (from $\theta_+$ to $\theta_-$ and vice versa)~\cite{Gralla:2019drh}. This parameter is primarily influenced by the BH mass, with significant corrections from the spin and the inclination angle for orientations that are near-edge-on combined with a high spin, especially for $\ang{60}\lesssim\theta_o\lesssim\ang{120}$ and $a\gtrsim 0.9M$.

Another critical parameter is the azimuthal angle variation within a half-libration, $\delta$~\cite{Gralla:2019drh}. It is more sensitive to the BH spin and inclination compared to the time delay. For face-on observation, it ranges from $\pi$ for a Schwarzschild BH to approximately $3\pi/2$ for $a/M = 0.99$~\cite{Gralla:2019drh}. A finite inclination angle, even as small as $\ang{17}$, can significantly deform $\delta$ at large spins, varying for photons entering bound orbits at different position angles. From Fig.~\ref{fig:images for various spins}, we see that the position angle shift of images with $N$ differing by $2$ becomes less regular as the spin or inclination angle increases. For low spin and nearly face-on observation, a simple relation for the position angle shift holds
\begin{equation}
\varphi^{N+2} - \varphi^N \approx 2\delta - 2\pi,
\end{equation}
offering an opportunity to measure the BH spin. We will discuss determining more general cases of BH parameters and source location in Sec.~\ref{sec:tomography}.

%%%%%%%%%%%%%%%%%%%%%%%%%%%%%%%%%%%%%
\subsection{Image Level Degeneracy}
\label{subsec:image degeneracy}
%%%%%%%%%%%%%%%%%%%%%%%%%%%%%%%%%%%%%

As illustrated in the previous subsection, there can be multiple images at the same level $N$ from a point source. This phenomenon does not occur for a Schwarzschild BH, where geodesics at each level lie entirely within the same plane containing the pointlike source, the BH, and the observer, sequentially projecting onto the image plane along a single line. In contrast, such cases become more common with increasing spin and inclination angle. This subsection delves into this phenomenon, termed image level degeneracy. 

In addition to the $7a, 7b$, and $7c$ level degeneracy in the example discussed in Sec.~\ref{subsubsec:example} and depicted in Fig.~\ref{fig:contour plots}, Fig.~\ref{fig:images}, and Table~\ref{tab:roots}, we examine the last panel of Fig.~\ref{fig:images for various spins}, which displays the most degenerate solutions. In this latter case, the source is located at $(r_s = 10\,M$, $\theta_s = \ang{90}$, $\phi_s = -\ang{45})$ outside a BH with spin $a = 0.8\,M$, reaching an observer at $(r_o = 1000\,M$, $\theta_o = \ang{80}$, $\phi_o = 0)$. To quantitatively understand the images, we present each image's polar direction sign $\nu_{\theta}^s$, polar half-orbit measure $n$, polar turning number $m$, and azimuthal winding number $k \equiv (\phi_f - \phi_o)/(2\pi)$ in Tables~\ref{tab:7abc details} and \ref{tab:80deg details}, respectively, for the two cases mentioned.

\begin{figure}[!htbp]
    \centering
    \includegraphics[width=0.4\textwidth]{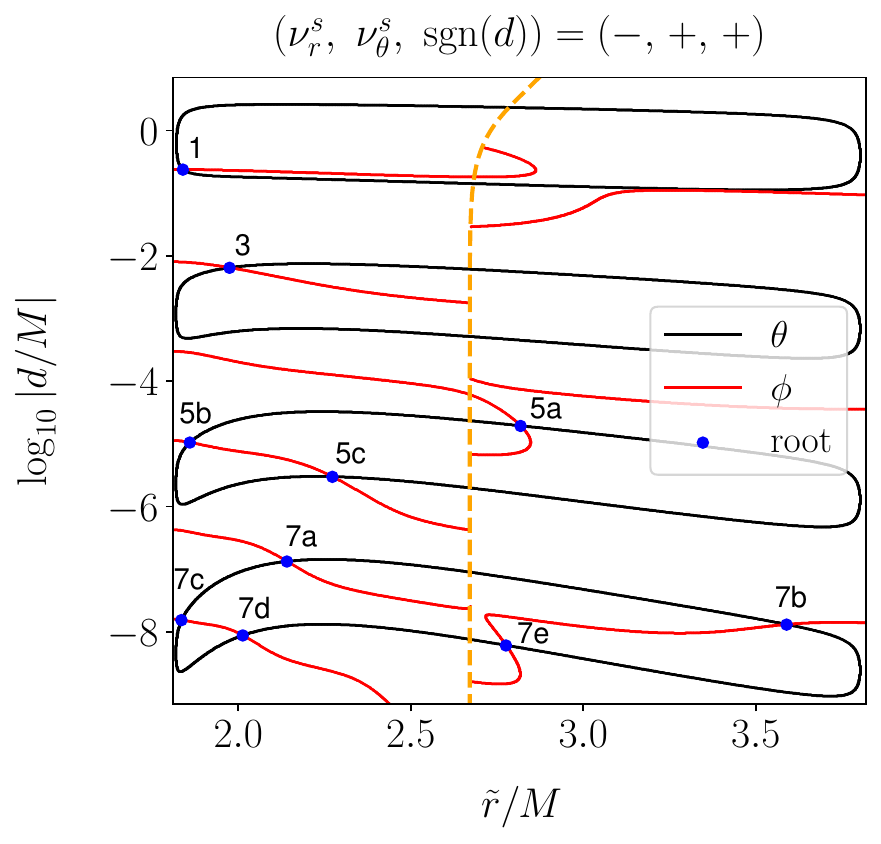}
    \includegraphics[width=0.4\textwidth]{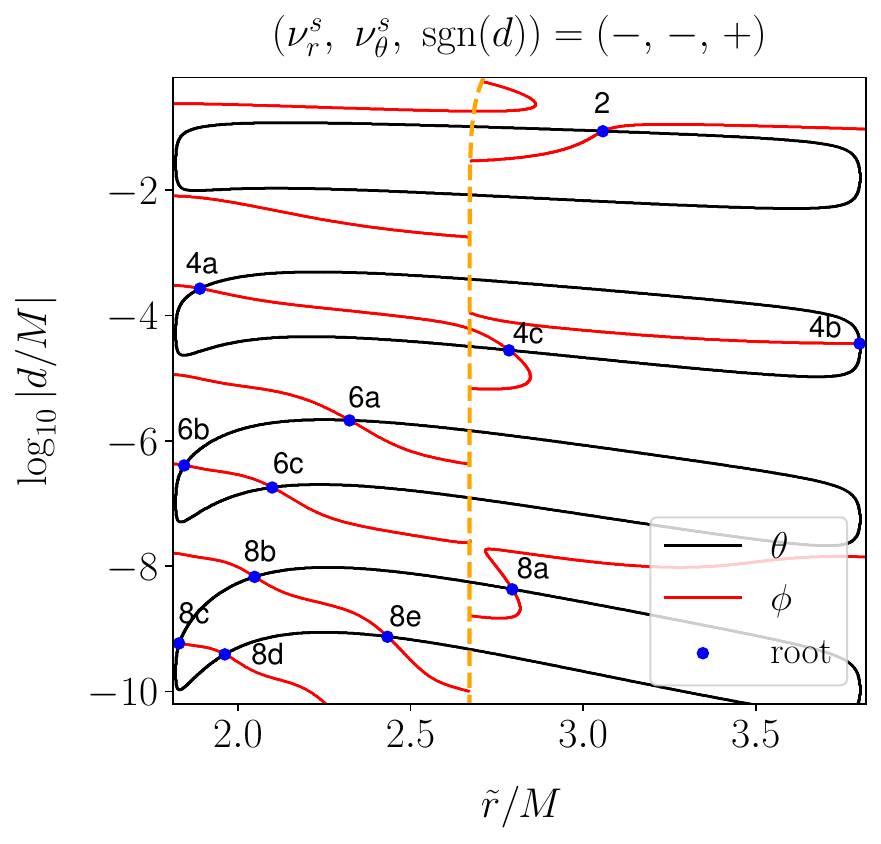}
    \caption{Contour plots used to find the lensed images for the example in the last panel of Fig.~\ref{fig:images for various spins}, following the same notation as Fig.~\ref{fig:contour plots}.}
    \label{fig:80deg contours}
\end{figure}

The corresponding contour plots of lensed photons are displayed in Fig.~\ref{fig:contour plots} and Fig.~\ref{fig:80deg contours} for each case. The polar contours for $\theta_f = \theta_o$ feature various closed loop lines. Given that the level number $N$ is defined in terms of half-orbits propagated in the $\theta$ direction, as specified in Eqs.~(\ref{eq:n}, \ref{eq:N}), each loop corresponds to one level. The upper and lower parts of each loop, however, differ by an additional turning in the $\theta$ direction. The azimuthal contour lines for $\phi_f = \phi_o + 2k\pi$ $(k \in \mathbb{Z})$ are distinguished by the winding number $k$. The yellow dashed lines, denoting the $\lambda = 0$ contour, separate regions for $k > 0$ and $k < 0$ among lensed photons. 

\begin{table}[!htbp]
\renewcommand\arraystretch{1.2} %% row spacing
\setlength\tabcolsep{7pt} %% column spacing
\centering
\begin{tabular}{ccccc}\hline\hline
Level & $\nu_{\theta}^s$ & $n$ & $m$ & $k$ \\ \hline
$7a$ & $+$ & 7.49 & 7 & $5$ \\
$7b$ & $+$ & 7.54 & 8 & $-3$ \\
$7c$ & $+$ & 7.59 & 8 & $-3$ \\ \hline
\end{tabular}
\caption{Detailed information on the angular motion of the degenerate images in Fig.~\ref{fig:contour plots}, Fig.~\ref{fig:images}, and Table~\ref{tab:roots}, including polar direction sign $\nu_\theta^s$, polar half-orbit measure $n$, polar turning number $m$, and azimuthal winding number $k \equiv (\phi_f - \phi_o)/(2\pi)$.}
\label{tab:7abc details}
\end{table}

\begin{table}[!htbp]
\renewcommand\arraystretch{1.2} %% row spacing
\setlength\tabcolsep{7pt} %% column spacing
\centering
\begin{tabular}{ccccc}\hline\hline

Level & $\nu_{\theta}^s$ & $n$ & $m$ & $k$ \\ \hline

$0$ & $-$ & 0.25 & 0 & $0$ \\

$1$ & $+$ & 1.82 & 2 & $1$ \\

$3$ & $+$ & 3.08 & 3 & $2$ \\

$5a$ & $+$ & 5.06 & 5 & $-2$ \\
$5b$ & $+$ & 5.14 & 5 & $4$ \\
$5c$ & $+$ & 5.94 & 6 & $4$ \\

$7a$ & $+$ & 7.07 & 7 & $5$ \\
$7b$ & $+$ & 7.09 & 7 & $-3$ \\
$7c$ & $+$ & 7.19 & 7 & $6$ \\
$7d$ & $+$ & 7.92 & 8 & $6$ \\
$7e$ & $+$ & 7.94 & 8 & $-3$ \\

$2$ & $-$ & 2.06 & 2 & $-1$ \\

$4a$ & $-$ & 4.11 & 4 & $3$ \\
$4b$ & $-$ & 4.44 & 4 & $-2$ \\
$4c$ & $-$ & 4.94 & 5 & $-2$ \\

$6a$ & $-$ & 6.06 & 6 & $4$ \\
$6b$ & $-$ & 6.16 & 6 & $5$ \\
$6c$ & $-$ & 6.93 & 7 & $5$ \\

$8a$ & $-$ & 8.06 & 8 & $-3$ \\
$8b$ & $-$ & 8.07 & 8 & $6$ \\
$8c$ & $-$ & 8.22 & 8 & $7$ \\
$8d$ & $-$ & 8.92 & 9 & $7$ \\
$8e$ & $-$ & 8.94 & 9 & $6$ \\ \hline
\end{tabular}
\caption{Detailed information on the angular motion of the images in the last panel of Fig.~\ref{fig:images for various spins}, following the same notation as Table~\ref{tab:7abc details}.}
\label{tab:80deg details}
\end{table}

Based on how the contour lines intersect in these examples, we categorize the origins of image level degeneracy as follows.

\begin{itemize}
    \item \textit{Extra polar turning}. 
    One azimuthal contour (red) intersects with a polar contour (black loop) twice, at both the upper and lower parts of the loop. The corresponding solutions differ by their $\theta$-direction turning numbers $m$ defined in Eq.~(\ref{eq:m}). For instance, in the last panel of Fig.~\ref{fig:images for various spins}, the images $5b$ and $5c$ begin with identical $\nu_{\theta}^s$ values but differ in their turning numbers $m$. They approach $\theta_o$ with decreasing and increasing $\theta$ propagation, respectively. For more details on these images and their corresponding contour plots, refer to Table~\ref{tab:80deg details} and Fig.~\ref{fig:80deg contours} (the same below).
    
    \item \textit{Opposite azimuthal motion}. 
    Two azimuthal contours (red) extending from the left and right sides intersect with one polar contour (black). Solutions in this category differ by the signs of their azimuthal momentum, characterized by opposite signs of $k$. For instance, the images $4a$ and $4b$ in the last panel of Fig.~{\ref{fig:images for various spins}}  originate from two distinct azimuthal contour lines.
    
    \item \textit{Extra azimuthal winding}. 
    Two azimuthal contours (red) extending from the same side intersect with a polar contour (black). This category includes cases with the same sign for $k$ but differing values. An example is images $6a$ and $6b$ in the last panel of Fig.~\ref{fig:images for various spins}.

    \item \textit{Combined case}. 
    A polar contour (black loop) intersects with one or more azimuthal contours (red) such that extra polar turnings coincide with either opposite azimuthal motions or additional azimuthal windings.
    
    \item \textit{Twin solutions}. 
    An azimuthal contour (red) intersects with a polar contour (black loop) twice at the same section (either upper or lower) of the loop. These solutions share the same initial $\theta$-momentum sign $\nu_{\theta}^s$ and turning number $m$, completing an identical number of azimuthal windings in the same direction. Consequently, they exhibit very similar values of $n$. An example is images $7b$ and $7c$ shown in Fig.~\ref{fig:contour plots}, Fig.~\ref{fig:images}, and Table~\ref{tab:roots}. 
\end{itemize}

%%%%%%%%%%%%%%%%%%%%%%%%%%%%%%%%%%%%%%%%%%%%%%%%%
%%%%%%%%%%%%%%%%%%%%
%\section{Distortion of Lensed Images.}
\section{Image Distortion in Kerr Spacetime}
\label{sec:distortion}
%%%%%%%%%%%%%%%%%%%%%%%%%%%%%%%%%%%%%%%%%%%%%%%%%
%%%%%%%%%%%%%%%%%%%%

The forward ray tracing method developed in Sec.~\ref{sec:FRLE} enables the calculation of geodesics that connect a specific point near the BH to a distant observer. In reality, emission sources, such as hotspots, have a finite volume. This section addresses this practical aspect by introducing a perturbative mapping technique~\cite{Cunningham1973,Ohanian1987}. This method relates small deviations near the original emission point to corresponding changes on the image plane. We formulate this perturbative mapping using conserved quantities to analyze the amplification both along and perpendicular to the critical curve for higher-order images. We then illustrate the scaling relations with an example in Schwarzschild spacetime, using straightforward geometric relations to calculate the amplification factor as a function of source position and image level.

%%%%%%%%%%%%%%%%%%%%%%%%%%%%%%%%%%%%%%%%%%%%%%%%%
\subsection{Perturbative Deviation of Forward Ray Tracing}
%%%%%%%%%%%%%%%%%%%%%%%%%%%%%%%%%%%%%%%%%%%%%%%%%

In our forward ray tracing approach, the final directions of a light ray, $(\theta_f,\,\phi_f)$, are determined by conserved quantities $\lambda$, $\eta$, momentum signs $\nu_r^s$, $\nu_{\theta}^s$, and the source position $\vec{r}_s = (r_s,\, \theta_s,\, \phi_s)$
\begin{equation}
\begin{aligned}
\theta_f&=\theta_f\left(\lambda,\,\eta,\,\nu_r^s,\,\nu_{\theta}^s \mid \vec{r}_s\right),\\
\phi_f&=\phi_f\left(\lambda,\,\eta,\,\nu_r^s,\,\nu_{\theta}^s \mid \vec{r}_s\right),
\end{aligned}
\end{equation}
where $\vec{r}_s$ specifies the starting point of the light ray, and the variables before the vertical line determine its initial direction.

In more realistic scenarios where the emission source occupies a finite volume centered around $\vec{r}_s$, the three-dimensional spatial extension of the source maps to a two-dimensional structure on each image. To account for this, we introduce a small displacement in the source position, $\delta \vec{r}_s \equiv (\delta r_s,\,\delta \theta_s,\,\delta \phi_s)$. For the light ray from this displaced point to still reach the observer at $(\theta_f,\,\phi_f)$, its initial direction must correspondingly change, characterized by deviations in the conserved quantities, $\delta \lambda$ and $\delta \eta$. The imaging condition (\ref{eq:imaging condition}) then leads to
\begin{equation}
\begin{aligned}
    0&=\delta\theta_{f}=\frac{\p \theta_{f}}{\p \lambda}\delta\lambda+\frac{\p \theta_{f}}{\p \eta}\delta\eta+\frac{\p \theta_{f}}{\p \vec{r}_s}\cdot\delta \vec{r}_s, \\ 
    0&=\delta\phi_{f}=\frac{\p \phi_{f}}{\p \lambda}\delta\lambda+\frac{\p \phi{_{f}}}{\p \eta}\delta\eta+\frac{\p \phi{_{f}}}{\p \vec{r}_s}\cdot\delta \vec{r}_s,
\end{aligned}
\end{equation}
where $\p/\p\vec{r}_s \equiv (\p/\p r_s, \p/\p \theta_s, \p/\p \phi_s)$. From this, we can solve $(\delta \lambda,\, \delta \eta)$ as a linear function of $\delta \vec{r}_s$.

We further translate the deviation into the image position using Eq.~(\ref{eq:impact parameters}). Combining these, we construct a perturbative map
\begin{equation}
    \left(\begin{matrix}
        \delta\alpha \\ 
        \delta\beta
    \end{matrix}\right) = \mathcal{M} \left(\begin{matrix}
        \delta r_s \\ 
        \delta \theta_s \\ 
        \delta \phi_s
    \end{matrix}\right),
\end{equation}
where the $2\times 3$ mapping matrix $\mathcal{M}$ is defined as
\begin{equation}
    \mathcal{M} \equiv - \begin{pmatrix}
        \dfrac{\p\alpha}{\p\lambda} & \dfrac{\p\alpha}{\p\eta} \\[1.1em]
        \dfrac{\p\beta}{\p\lambda} & \dfrac{\p\beta}{\p\eta}
    \end{pmatrix}
    \begin{pmatrix}
        \dfrac{\p\theta_{f}}{\p\lambda} & \dfrac{\p\theta_{f}}{\p\eta} \\[1.1em]
        \dfrac{\p\phi_{f}}{\p\lambda} & \dfrac{\p\phi_{f}}{\p\eta}
    \end{pmatrix}^{-1}
    \begin{pmatrix}
        \dfrac{\p\theta_{f}}{\p \vec{r}_s} \\[1.1em] 
        \dfrac{\p\phi_{f}}{\p \vec{r}_s} 
    \end{pmatrix}.
    \label{eq:shape matrix}
\end{equation}
Here, $\p\theta_f/\p\vec{r}_s$ and $\p\phi_f/\p\vec{r}_s$ both represent row vectors, following our definition of $\partial/\partial \vec{r}_s$. Utilizing the definitions of $\alpha$ and $\beta$ in Eq.~(\ref{eq:impact parameters}), along with the integral formulations for $\theta_f$ in Eq.~(\ref{eq:theta_f}) and $\phi_f$ in Eqs.~(\ref{eq:integral form},\,\ref{eq:path integrals}), the mapping matrix can be evaluated. For practical examples of these calculations, refer to the \textit{Mathematica} notebook provided in \cite{KerrP2P}.

\begin{figure}[t]
    \centering
    \includegraphics[width=0.45\textwidth]{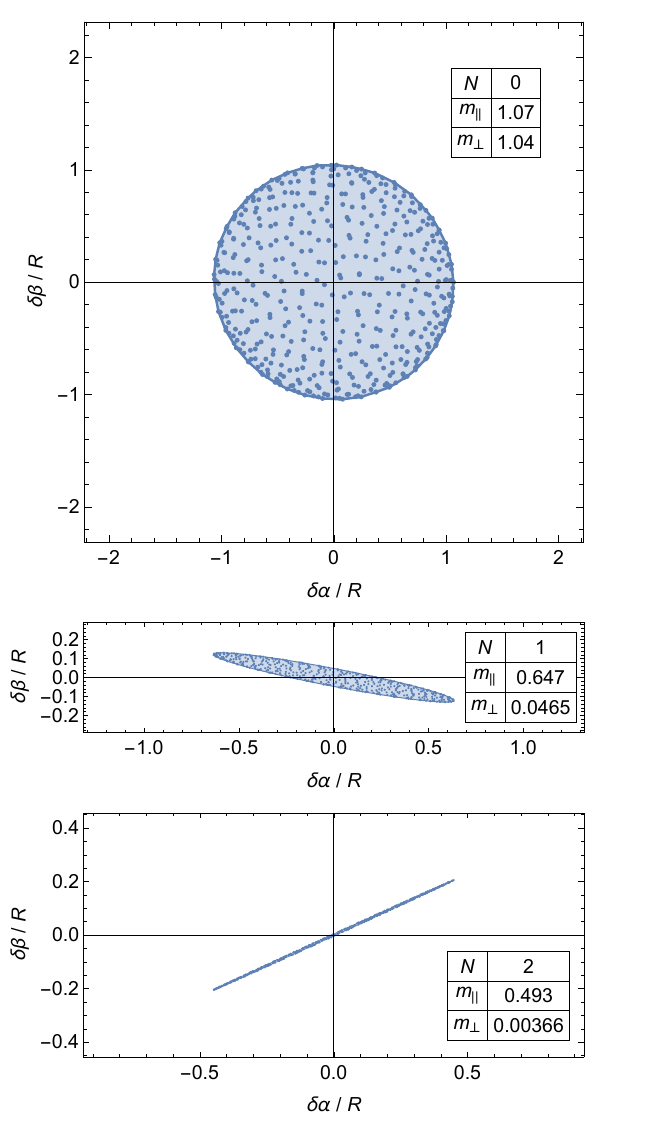}
    \caption{Images of levels $N=0, 1$, and $2$ for a spherical emission source with radius $R$, centered at the emission point described in Sec.~\ref{subsubsec:example}, using the same BH parameters. The coordinate origins on the image plane correspond to the image position of the emission center. Sample points within the spherical source are randomly selected, each yielding a blue point on the image plane within an elliptical envelope. The amplification rates, $m_{\parallel}$ and $m_{\perp}$, defined as the ratios of the major and minor axes of these ellipses to $2R$, respectively, are displayed for each image level.} 
    \label{fig:shape}
\end{figure}

\begin{figure}[t]
    \centering
    \includegraphics[width=0.45\textwidth]{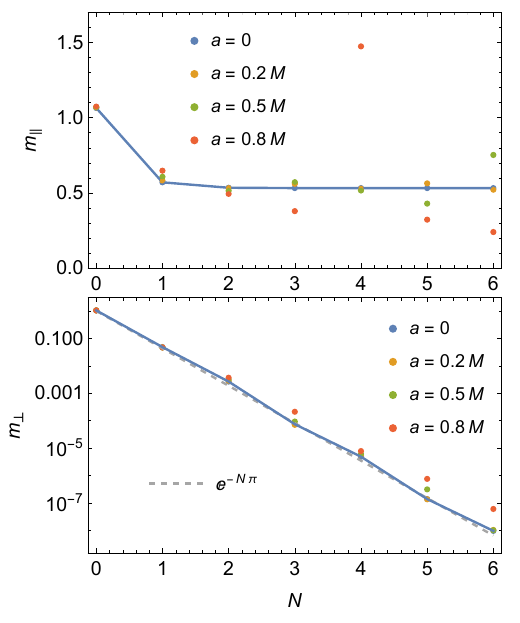}
    \caption{Amplification rates $m_{\parallel}$ (top) and $m_{\perp}$ (bottom) for different image levels $N$, plotted for the same scenario as in Fig.~\ref{fig:shape} but with varying BH spins. These rates, defined as the ratios of the ellipse axes parallel and perpendicular to the critical curve to  spherical source diameter ($2R$), show how image distortion varies with spin and image level. The blue solid lines represent the Schwarzschild case ($a=0$), while colored dots indicate different spin values. The gray dashed line illustrates the exponential scaling $\propto \ee^{-N\pi}$.}
    \label{fig:mag vs. N}
\end{figure}

\begin{figure*}
    \centering
    \includegraphics[width=0.95\textwidth]{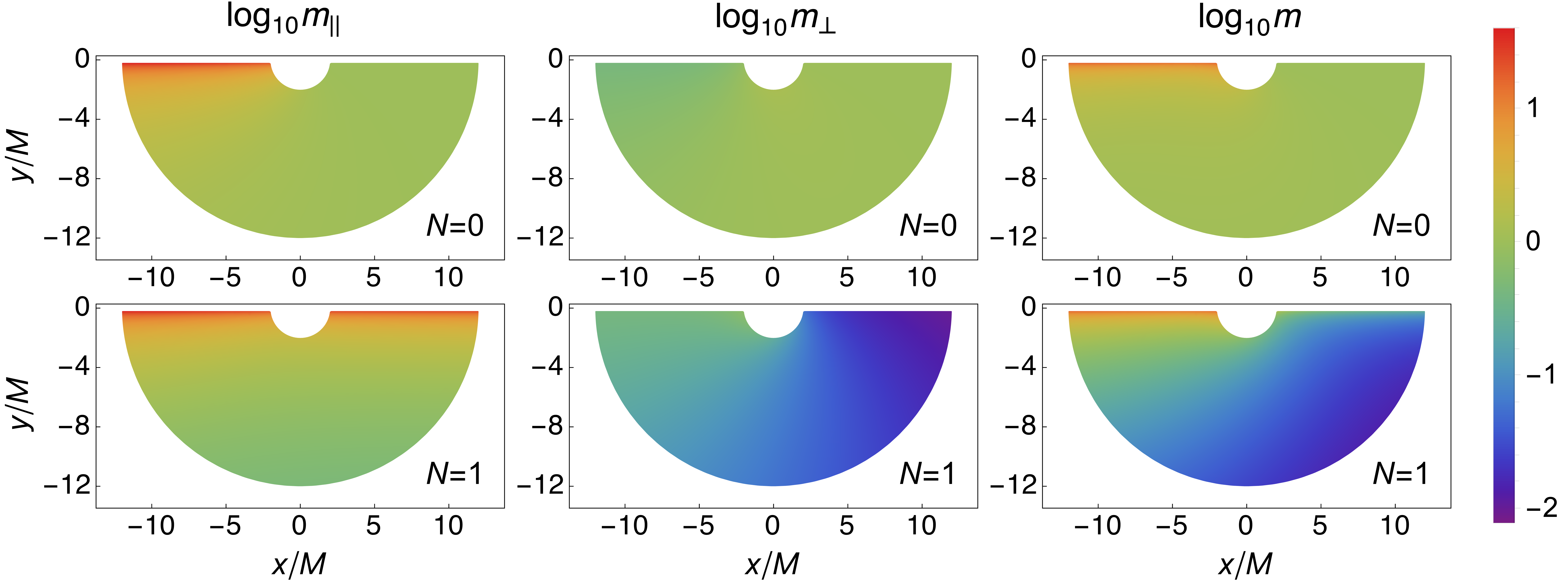}
    \caption{Amplification rates $m_{\parallel}$ and $m_{\perp}$, parallel and perpendicular to the critical curve, respectively, along with the area amplification rate $m \equiv m_{\parallel}m_{\perp}$, for $N=0$ and $N=1$ level images of a point source outside a Schwarzschild BH. The geodesics are confined to the $x$-$y$ plane at $\theta = \pi/2$, with the observer positioned at $(x=1000\,M, y=0)$. For illustrative purposes, a constraint of $y < -0.2\,M$ is imposed to avoid divergence at the caustics ($y=0$).}
    \label{fig:mag distribution}
\end{figure*}

We consider an emission source with a finite spherical volume of radius $R$, with the spacetime configuration regarding the source center, the BH, and the observer the same as that described in Sec.~\ref{subsubsec:example}. To ensure the perturbative mapping remains valid, the spherical radius $R$ must be much smaller than $M$. More complex shapes of emission sources can be approximated by aggregating multiple such spherical volumes. The shapes of the first three image levels ($N = 0, 1, 2$) are depicted in Fig.~\ref{fig:shape}. By comparing the shapes of lensed images ($N = 1$ and $N = 2$) to their positions on the image plane, as depicted in Fig.~\ref{fig:images}, it is evident that these images typically manifest as ellipses, with the major axis aligning parallel to the critical curve and the minor axis significantly compressed in the perpendicular direction. The amplification rates, $m_{\parallel}$ and $m_{\perp}$, are defined respectively as the ratios of the ellipse's axes—parallel and perpendicular to the critical curve—to the source diameter $2R$. These rates are determined by the absolute values of the two singular values of the mapping matrix $\mathcal{M}$. The amplification rates for each image level are displayed in the figure.

We further explore higher-order images and the effects of BH spin, illustrating the corresponding amplification rates $m_{\parallel}$ and $m_{\perp}$ in Fig.~\ref{fig:mag vs. N}. The blue lines denote the Schwarzschild case with $a = 0$. Although there are deviations at high spins, the overall trend in amplification rates still adheres to that observed in the Schwarzschild scenario. In the following subsection, we will derive these scaling relations using straightforward geometric relations, showing that $m_{\parallel}$ remains relatively stable at approximately $\mathcal{O}(M/\rho_S)$, where $\rho_S = r_s \left|\sin(\phi_o - \phi_s)\right|$, across higher levels, while $m_{\perp}$ decreases exponentially as $\propto \ee^{-N\gamma} \approx \ee^{-N\pi}$.

Spatial mapping plays a vital role in observing hotspots since the total intensity of each image is roughly proportional to its area, a consequence of Liouville's theorem~\cite{Misner:1973prb}. In Fig.~\ref{fig:mag distribution}, we demonstrate how $m_{\parallel}, m_{\perp}$, and the area amplification rates, $m \equiv m_{\parallel}m_{\perp}$, vary depending on the source's position relative to both the BH and the observer in Schwarzschild spacetime for image levels $N=0$ and $N=1$. Typically, higher-order images are exponentially dimmer, thereby posing observational challenges. Nonetheless, sources positioned near caustics can yield exceptionally amplified images.

Additionally, the duration of emission is a crucial observational parameter. For the spherical emission model considered here, we analyze the depth in the time domain by tracking and comparing the distribution of arrival times $t_f$ for photons emitted from the sphere at a fixed moment. Our findings confirm that the time-domain depth of each image approximately equals the coordinate time it takes a photon to travel across the source diameter, $2R$.

\subsection{Image Distortion in Schwarzschild Spacetime}
\label{subsec:Sch distortion}
This subsection derives the amplification rate for a hotspot outside a Schwarzschild BH using straightforward geometric relations, similar to the approach in Ref.~\cite{Ohanian1987}. We assume, without loss of generality, that the orbital plane containing the hotspot's center, the BH, and the observer coincides with the equatorial plane at $\theta = \pi/2$. Under these conditions, the formalism in Sec.~\ref{subsec:Trajectory} reduces to $p^{\theta} = 0$, $\eta = 0$, $\mathcal{R}(r) = r^4 - \lambda^2 r^2(1 - 2M/r)$, and $\Theta(\theta) = 0$. To bypass the apparent singularity in the angular integrals, the orbital motion is approximated as the limit of $\eta \to 0^+$, which entails infinitesimal polar oscillations between $\theta_{\mp} = \pi/2\,\mp\,\epsilon$, with $\epsilon \to 0$. This approximation simplifies the angular integrals in Eqs.~(\ref{eq:inteq1}, \ref{eq:Gtheta}, \ref{eq:Gphi}) to $G_{\theta} = G_{\phi} = \tau$. This value is derived from Eqs.~(\ref{eq:inteq1}, \ref{eq:Ir}) as follows
\begin{equation}
    \tau=\fint_{r_s}^{r_f}\frac{\nu_r\dd r}{\sqrt{\R(r)}}=\fint_{r_s}^{r_f}\frac{\nu_r\dd r}{\sqrt{r^4-\lambda^2r^2(1-2M/r)}}.\label{eq:tauS}
\end{equation}
Accordingly, with $I_{\phi} = 0$ as defined in Schwarzschild spacetime (see Eq.~(\ref{eq:Iphi})), the final azimuthal angle of the photon as expressed in Eq.~(\ref{eq:inteq2}) becomes 
\begin{equation}
\phi_f = \phi_s + \lambda \tau.\label{eq:phifS}
\end{equation}
Thus, the condition for successful forward ray tracing reduces to $\phi_f = \phi_o + 2k\pi$ (where $k \in \mathbb{Z}$) compared to Eq.~(\ref{eq:imaging condition}).

\begin{figure}[!htbp]
    \centering
    \includegraphics[width=0.45\textwidth]{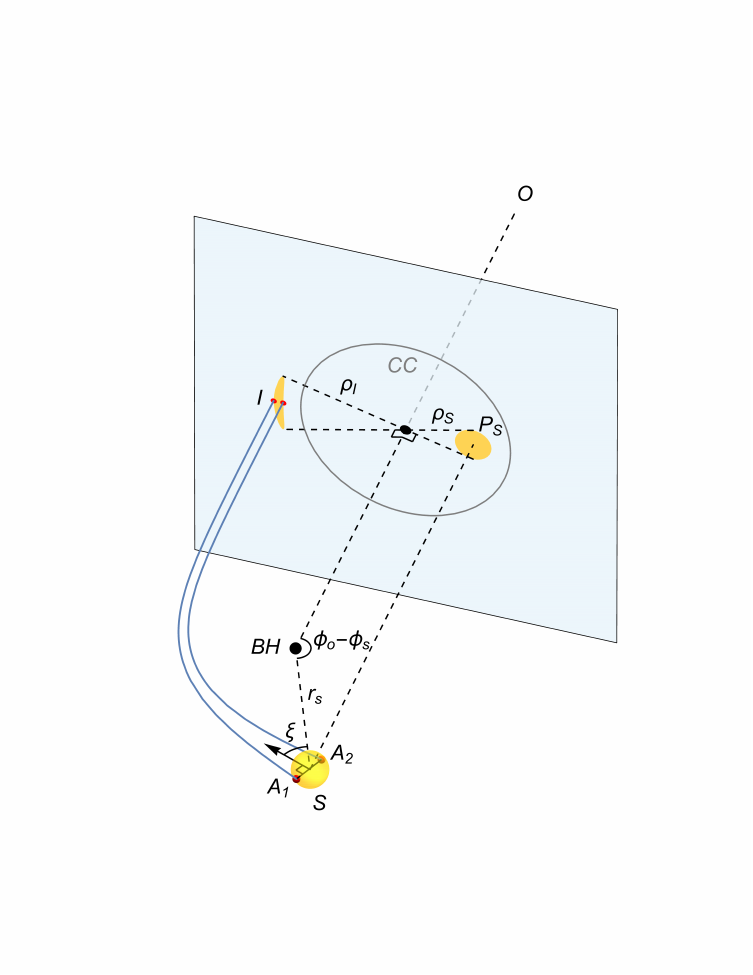}
    \caption{Illustration of geometric relations for a spherical emission source ($S$) and its $N=1$ level image ($I$) on the image plane. $\rho_I$ and $\rho_S$ denote the distances from the BH projection on the image plane to $I$ and the projection of $S$ ($P_S$), respectively. The source center's orbital plane aligns with the BH equatorial plane ($\theta=\pi/2$), and is located at $(r_s, \phi_s)$. The initial direction of geodesics from the source center is shown by the black arrow, forming an angle $\xi$ with the direction toward the BH. The two image edges perpendicular to the critical curve (CC) are marked by red dots, corresponding to points $A_1$ and $A_2$ on the spherical source.}
    \label{fig:3Dmag}
\end{figure}

To clearly illustrate the derivation of the amplification rate, we present the geometric relations for a radiating spherical hotspot in Fig.~\ref{fig:3Dmag}. The hotspot, denoted by $S$, has its center at $(r_s,\,\theta_s = 0,\,\phi_s)$, with the initial direction of geodesics from this point indicated by the black arrow, leading to an image $I$ on the image plane. Although the example shown uses geodesics for $N=1$, the explanation applies to all image levels. The direct projection of the black hole on the image plane serves as the origin of the coordinate system, marked by a black dot, and the projection of the hotspot's center is labeled as  $P_S$. $\rho_I = \sqrt{\alpha^2 + \beta^2}$ and $\rho_S$ represent the distances from the origin to $I$  and $P_S$, respectively. The gray contour line labeled CC on the image plane denotes the critical curve.

Each point within the spherical hotspot forms an orbital plane, which can be different from the one containing the source center. We use dashed lines on the image plane to show the boundary of the region containing intersecting lines between orbital planes and the image plane. From this, we can easily determine the amplification rate along the critical curve
\begin{equation}
    m_{\parallel}=\frac{\rho_I}{\rho_{S}}=\left|\frac{\lambda}{r_s\sin(\phi_o-\phi_s)}\right|,
    \label{eq:mpar}
\end{equation}
assuming the hotspot is small enough, i.e., $R \ll M$. Here, the relation $\rho_I = \left|\lambda\right|$ follows from Eq.~(\ref{eq:impact parameters}) taking $\theta_o = \pi/2$ and $\eta = 0$, and $\rho_S = r_s \left|\sin(\phi_o - \phi_s)\right|$ follows from simple geometric relations. Each image level is characterized by $\lambda$, the absolute value of which asymptotically approaching the critical value $|\tilde{\lambda}| = 3\sqrt{3}\,M$~\cite{Luminet:1979nyg} as $N$ becomes large in the Schwarzschild case. 

Notice that Eq.~(\ref{eq:mpar}) becomes divergent as $\left(\phi_o - \phi_s\right)$ approaches an integer multiple of $\pi$, marking instances where the source is collinear with the BH and the observer, along a line known as the caustic. However, in practical scenarios with a finite-sized source, this divergence in the amplification rate is regulated. Specifically, near the alignment where $\rho_S = r_s |\sin(\phi_o - \phi_s)| \approx 0$, the apparent divergence of $1/\rho_S$ in Eq.~(\ref{eq:mpar}) is compensated by the area integral of the source projection, $\propto \rho_S \dd \rho_S$, resulting in a finite contribution to the overall intensity. To clarify, consider a source with radius $R$ positioned directly on the caustic. The overall amplification rate, defined as the ratio of the total image area to the source's projected area and calculated as a weighted average across the source projection, is expressed as
\begin{equation}
\begin{aligned}
m_{\text{overall}} &\equiv \frac{\int_0^R m_{\perp} m_{\parallel} \, 2\pi\rho_S \, \dd\rho_S}{\pi R^2} \\
&\simeq \frac{m_{\perp}R \times 2\pi\rho_I}{\pi R^2} \\ 
&= m_{\perp} \frac{2\rho_I}{R}.
\end{aligned}
\end{equation}
Thus, the overall amplification rate is moderated by  the ratio $\rho_I/R$, accounting for the source's finite size.

We next consider the amplification rate in the direction perpendicular to the critical curve, $m_{\perp}$. The angle between the initial direction of the geodesics from $S$ and the direction from the BH to the source is defined as $\xi$, following the geometric relation
\begin{equation}
    \tan\xi=\frac{r p^{\phi}}{p^{r}}\bigg\rvert_{S}=\nu_r^s\operatorname{sgn}(\lambda)\frac{1}{\sqrt{r_s^2/\lambda^2-1+2M/r_s}},
    \label{eq:xi}
\end{equation}
where the second equality can be derived using Eqs.~(\ref{eq:pr}, \ref{eq:pphi}). Here, $\lambda$ is the conserved quantity carried by the light ray from $S$.

We focus on the two edges of the image, marked in red dots on the image plane in Fig.~\ref{fig:3Dmag}, originating from the two points $A_1$ and $A_2$ on the spherical source, whose connecting line is perpendicular to the initial direction. The two geodesics, shown in blue lines, are localized on the BH equatorial plane. The coordinates of $A_1$ and $A_2$ can be parameterized as
\begin{equation}
\begin{aligned}
        \vec{r}_{A_1} &= (r_s+\delta r,\ \pi/2,\ \phi_s+\delta\phi),\\
        \vec{r}_{A_2} &= (r_s-\delta r,\ \pi/2,\ \phi_s-\delta\phi),
\end{aligned}
\label{eq:rA1A2}
\end{equation}
where $\delta r \equiv R \sin \xi$ and $\delta \phi \equiv -R \cos \xi / r_s$. We characterize their conserved quantities as $\lambda \pm \delta \lambda$, respectively. The width of the image perpendicular to the critical curve is $2|\delta\lambda|$, making the amplification rate $m_\perp = |\delta\lambda / R|$.

The azimuthal part of the condition in Eq.~(\ref{eq:imaging condition}) leads to
\begin{equation}
   0=\delta\phi_{f} = \frac{\p\phi_f}{\p\lambda} \delta\lambda + \frac{\p\phi_f}{\p{r_s}} \delta r_s + \frac{\p\phi_f}{\p{\phi_s}} \delta\phi_s.
    \label{eq:Sch perturbation}
\end{equation}
We consider a special case where the variations in $r_s$ and $\phi_s$ align with the original geodesic of the hotspot center, implying $\delta r_s/\delta \phi_s = p^r/p^\phi$. Consequently, geodesics originating from this new position retain the same $\lambda$ as the original, i.e., $\delta \lambda = 0$, leading to the relation
\begin{equation}
    \frac{\p\phi_f}{\p r_s} + \frac{\p\phi_f}{\p \phi_s}\frac{\tan\xi}{r_s} = 0,\label{eq:phifsimp}
\end{equation}
where the relation in Eq.~(\ref{eq:xi}) was used. Next, we consider the spatial displacement corresponding to the two edge points, $A_1$ and $A_2$. By incorporating Eq.~(\ref{eq:phifsimp}) into Eq.~(\ref{eq:Sch perturbation}) and setting $\delta r_s = \delta r$ and $\delta \phi_s = \delta \phi$, as defined in Eq.~(\ref{eq:rA1A2}), and using the relation $\partial \phi_f / \partial \phi_s = 1$ as suggested by Eq.~(\ref{eq:phifS}), we obtain
\begin{equation}
    \delta\lambda=\frac{R}{r_s\cos\xi}\left(\frac{\p\phi_f}{\p\lambda}\right)^{-1}.
\end{equation}
Here, $\delta \lambda$ represents the width of the image $I$ perpendicular to the critical curve, leading to the amplification rate
\begin{equation}
    m_{\perp}=\left|\frac{1}{r_s\cos\xi}\left(\frac{\p\phi_f}{\p \lambda}\right)^{-1}\right|,
    \label{eq:mperp}
\end{equation}
which can be further elaborated using the expressions provided in Eqs.~(\ref{eq:tauS}, \ref{eq:phifS}).

We examine the asymptotic behavior of $m_\perp$ for higher image levels. Near the critical curve, the total azimuthal winding $\phi_f - \phi_s$ can be approximated as~\cite{Luminet:1979nyg,Gralla:2019xty,Gralla:2019drh,Tsukamoto:2020iez}
\begin{equation}
    |\phi_f - \phi_s| \sim \ln\frac{C}{|\lambda-\tilde{\lambda}|},\qquad \lambda\to \tilde{\lambda},
\label{eq:phifphis}
\end{equation}
where $C$ is a numerical factor largely dependent on the radial position of the source, $r_s$, which determines the path type experienced by the photon. For example, a source far from the BH results in $C \approx 80.6\,M$, while a source just outside the horizon results in $C \approx 21.6\,M$~\cite{Gralla:2019xty}. Additionally, the azimuthal winding increases approximately with image level $N$ as $|\phi_f - \phi_s| \sim N\,\pi$. Integrating this with Eq.~(\ref{eq:phifphis}) suggests
\begin{equation}
    \left|\frac{\p\phi_f}{\p\lambda}\right| \sim \frac{1}{|\lambda-\tilde{\lambda}|}\propto \frac{\ee^{N\pi}}{C}.
\end{equation}
Applying this to Eq.~(\ref{eq:mperp}) yields the asymptotic scaling relation
\begin{equation}
    m_{\perp}\propto\frac{C\,\ee^{-N\,\pi}}{r_s|\cos\xi|},
\end{equation}
demonstrating that as $N$ increases, higher-order images are exponentially compressed in the direction perpendicular to the critical curve. This relation also highlights that the compression perpendicular to the critical curve is intensified by a large $r_s$.

%%%%%%%%%%%%%%%%%%%%%%%%%%%%%%%%%%%%%%%%%%%%%%%%%
\section{Spacetime Tomography and Hotspot Localization}
\label{sec:tomography}
%%%%%%%%%%%%%%%%%%%%%%%%%%%%%%%%%%%%%%%%%%%%%%%%%
The forward ray tracing method is adept at simulating images of hotspots. This section demonstrates how this method can be applied to derive observational data, specifically for measuring the BH mass, spin, inclination angle, and hotspot location based on the identification and analysis of both the direct ($N=0$) and the first lensed ($N=1$) images of a hotspot. Since the intensity of higher-order images is exponentially suppressed, our focus will primarily be on these two initial images.

Emissions near a BH can exhibit significant time variability, which may be reflected across various image levels. The intensity of these images incorporates this variability, albeit with specific time delays. Thus, by correlating intensity fluctuations across different regions on the image plane with corresponding time delays, we can confirm their common origin as emissions from the same source~\cite{Hadar:2020fda}. The time delay between the two lowest image levels from a pointlike emission is sensitive to the BH mass, while their positional angle shift along the photon ring is sensitive to the BH spin~\cite{Gralla:2019drh,Hadar:2020fda,Andrianov:2022snn}. Additionally, the observer's inclination angle to the BH and the hotspot's spatial location can influence both the time delay and position angle shift, complicating the analysis as demonstrated in Sec.~\ref{subsec:delay and shift}. Forward ray tracing can mitigate these complexities by facilitating the scan over various source locations.

 We model a pointlike source located at $(r_s,\theta_s,\phi_s)$ outside a Kerr BH with mass $M$ and spin $a$, directing sequential geodesics toward an observer positioned at an inclination angle $\theta_o$ relative to the BH. Without loss of generality, we set the observer's azimuthal angle $\phi_o = 0$. By successfully detecting both the direct emission ($N=0$) and the first lensed image ($N=1$) of a hotspot through correlated time variability of intensity, we can ascertain precise locations of these images on the image plane, represented by
\begin{equation}
    (\alpha_0, \beta_0),\quad (\alpha_1, \beta_1),\label{eq:2levels}
\end{equation}
Additionally, the time difference between their arrivals can be quantified as
\begin{equation}
    \Delta t \equiv t_1 - t_0,\label{eq:2levelst}
\end{equation}
where $t_0$ and $t_1$ denote the arrival times of photons emitted simultaneously from the hotspot for $N=0$ and $N=1$ image levels, respectively. 

Notice that the identification of locations on the image plane in Eq.~(\ref{eq:2levels}) requires calibration of the image plane. This calibration includes: {(\romannumeral1)} translating angular distances into units of length, dependent on the distance between the BH and the observer; {(\romannumeral2)} establishing the coordinate origin at the projection of the BH; and {(\romannumeral3)} defining the spin projection axis $\beta$, as described below Eq.~(\ref{eq:impact parameters}). The distance is well-measured for astrophysical BHs, thus not an issue for our analysis. For a nearly face-on observer, with $\theta_o$ close to $0$ or $\pi$, the critical curve appears almost circular, and the BH's projection aligns closely with its geometric center. Consequently, the uncertainty in determining the coordinate origin is negligible compared to the uncertainties in the image locations, making it not a concern for our purposes. However, identifying the spin projection requires observing a jet from the BH, which may not always be present. Therefore, we consider the position angle (PA) of the spin projection on the image plane as an unknown parameter to be determined when a jet is not observable.

\begin{table}[!htbp]
    \centering
    \renewcommand\arraystretch{1.3} %% row spacing
    \setlength\tabcolsep{3pt} %% column spacing
    \begin{tabular}{cc}\hline\hline
         BH and Observer & Source \\ \hline
        mass $M$  & \multirow{4}{*}{\makecell[c]{location \\ $(r_s,\theta_s,\phi_s)$}} \\ 
        spin $a$ & \\ 
        inclination angle $\theta_o$ & \\ 
        PA of spin projection & \\ \hline
    \end{tabular}
    \caption{Parameters to be determined through observation of $N=0$ and $N=1$ images from a hotspot, including both BH and observer-related quantities as well as source location. `PA' refers to the position angle.}
    \label{tab:tomoparameters}
\end{table}

We summarize all relevant parameters to be determined in Table~\ref{tab:tomoparameters}, encompassing the quantities related to the BH and the observer, as well as the location of the source. We will explore how observational data listed in (\ref{eq:2levels}, \ref{eq:2levelst}) can be utilized to ascertain these parameters. Our discussion will bifurcate into two scenarios, contingent upon the extent of prior knowledge we possess about the target SMBHs observed by EHT and ngEHT.

The variables listed in Table~\ref{tab:tomoparameters} can be estimated using Monte Carlo Markov chain (MCMC) simulations. We define the set of random variables as
\begin{equation}
\bm{\Theta} \equiv \left(M,\ a,\ \theta_o,\ \text{PA},\ r_s^{(j)},\ \theta_s^{(j)},\ \phi_s^{(j)},\ \cdots \right)
\label{eq:Theta}
\end{equation}
and the observables from hotspots as 
\begin{equation}
\bm{X}^{(j)} \equiv \left(\alpha_0^{(j)},\ \beta_0^{(j)},\ \alpha_1^{(j)},\ \beta_1^{(j)},\ \Delta t^{(j)} \right),
\label{eq:Xj}
\end{equation}
where $j=a,b,c,\ldots$ labels the $j$th hotspot observed. Each selection of $\bm{\Theta}$ generates pairs of direct and the first lensed images $\bm{X}^{(j)}(\bm{\Theta})$ through the forward ray tracing method. These are then compared with the observed values $\overbar{\bm{X}^{(j)}}$, which in the following two examples are assumed to match the theoretical values computed from the true parameters. Consequently, we introduce the likelihood function
\begin{equation}
    p\left(\bm{X}\middle|\bm{\Theta}\right) = \prod_{k,j}\frac{1}{\sqrt{2\pi}\sigma_{\bm{X}_k^{(j)}}}\exp\left[-\frac{(\bm{X}^{(j)}_k(\bm{\Theta})-\overbar{\bm{X}^{(j)}_k})^2}{2\sigma_{\bm{X}_k^{(j)}}^{2}}\right],
\end{equation}
where $k$ labels each component of the vector $\bm{X}^{(j)}$, and $\sigma_{\bm{X}_k^{(j)}}$ corresponds to its measurement uncertainty.  

In scenarios where some variables in $\bm{\Theta}$ are predetermined by other measurements, such as the inclination angle $\theta_o$, the spin projection PA, and the SMBH mass $M$, we employ the prior probability function
\begin{equation}
    p(\bm{\Theta}) = \prod_{l} \frac{1}{\sqrt{2\pi}\sigma_{\bm{\Theta}_l}}\exp\left[-\frac{(\bm{\Theta}_l-\overbar{\bm{\Theta}_l})^2}{2\sigma_{\bm{\Theta}_l}^2}\right],
\end{equation}
where $l$ labels the predetermined parameters with true values $\overbar{\bm{\Theta}_l}$ and uncertainties $\sigma_{\bm{\Theta}_l}$. According to Bayes' theorem, the posterior probability is then
$p\left(\bm{\Theta}\middle|\bm{X}\right)\propto p\left(\bm{X}\middle|\bm{\Theta}\right) p(\bm{\Theta})$. 

The Python package \textit{emcee}~\cite{Foreman_Mackey_2013} is utilized for the MCMC simulation. During the simulation, we allow $\bm{\Theta}$ to vary, starting from their true values. The distribution of the resulting images and time delays are then used to compute the posterior probability. We fix the observer distance at $r_o=1000\,M$, where the results have converged to those at larger distances.

In the following two subsections, we analyze the two target SMBHs, M87$^*$~\cite{EventHorizonTelescope:2019ggy} and Sgr A$^*$~\cite{EventHorizonTelescope:2022exc}, where the detection of hotspots around is feasible with the current capabilities of the EHT and is anticipated to be more promising with its next-generation upgrade.

\subsection{M87$^*$ with Predetermined PA and $\theta_o$}
\label{subsec:M87}

Observation of the M87$^*$ jet have established the PA of the spin projection and $\theta_o = {(163\pm 2)}^\circ$~\cite{EventHorizonTelescope:2019pgp}. Consequently, there remain $5$ parameters to determine in Table~\ref{tab:tomoparameters}, including the BH mass and spin, and the hotspot location, assuming a single hotspot observation. This number of unknowns matches the number of observables provided in Eq.~(\ref{eq:Xj}) for the two lowest level images from a single hotspot, thus allowing a successful correlation observation of the hotspot to potentially determine all relevant parameters.

Consider the example of a BH with a mass $\overbar{M}$ and a spin $\overbar{a/M} = 0.8$, a hotspot at $(\overbar{r_s},\,\overbar{\theta_s},\,\overbar{\phi_s}) = (10\,\overbar{M},\,\ang{90},\ \ang{-45})$, observed from $\phi_o = 0$ and $\overbar{\theta_o} = \ang{163}$ with $\sigma_{\theta_o} = \ang{2}$. Using forward ray tracing, the corresponding true values of impact parameters for emissions $N=0$ and $N=1$ are $(\overbar{\alpha_0},\,\overbar{\beta_0}) = (-7.45\,\overbar{M},\, 7.32\,\overbar{M})$ and $(\overbar{\alpha_1},\,\overbar{\beta_1}) = (1.62\,\overbar{M},\, -5.30\,\overbar{M})$, respectively, with a time delay $\overbar{\Delta t} = 29.57\,\overbar{M}$. As previously mentioned, the central values of the observables are set to match their theoretical true values. We assume measurement uncertainties for $N=0$ and $N=1$ as $\sigma_{\alpha_0} = \sigma_{\beta_0} = \overbar{M}$ and $\sigma_{\alpha_1} = \sigma_{\beta_1} = 2\,\overbar{M}$, reflecting the angular resolution expected from the ngEHT~\cite{Chael:2021rjo,Lico:2023mus}, with a time resolution $\sigma_{\Delta t} = 0.5\,\overbar{M}$. These parameters and uncertainties are listed in Table~\ref{tab:set M87}. 

\begin{figure*}
    \centering
    \includegraphics[width=0.9\textwidth]{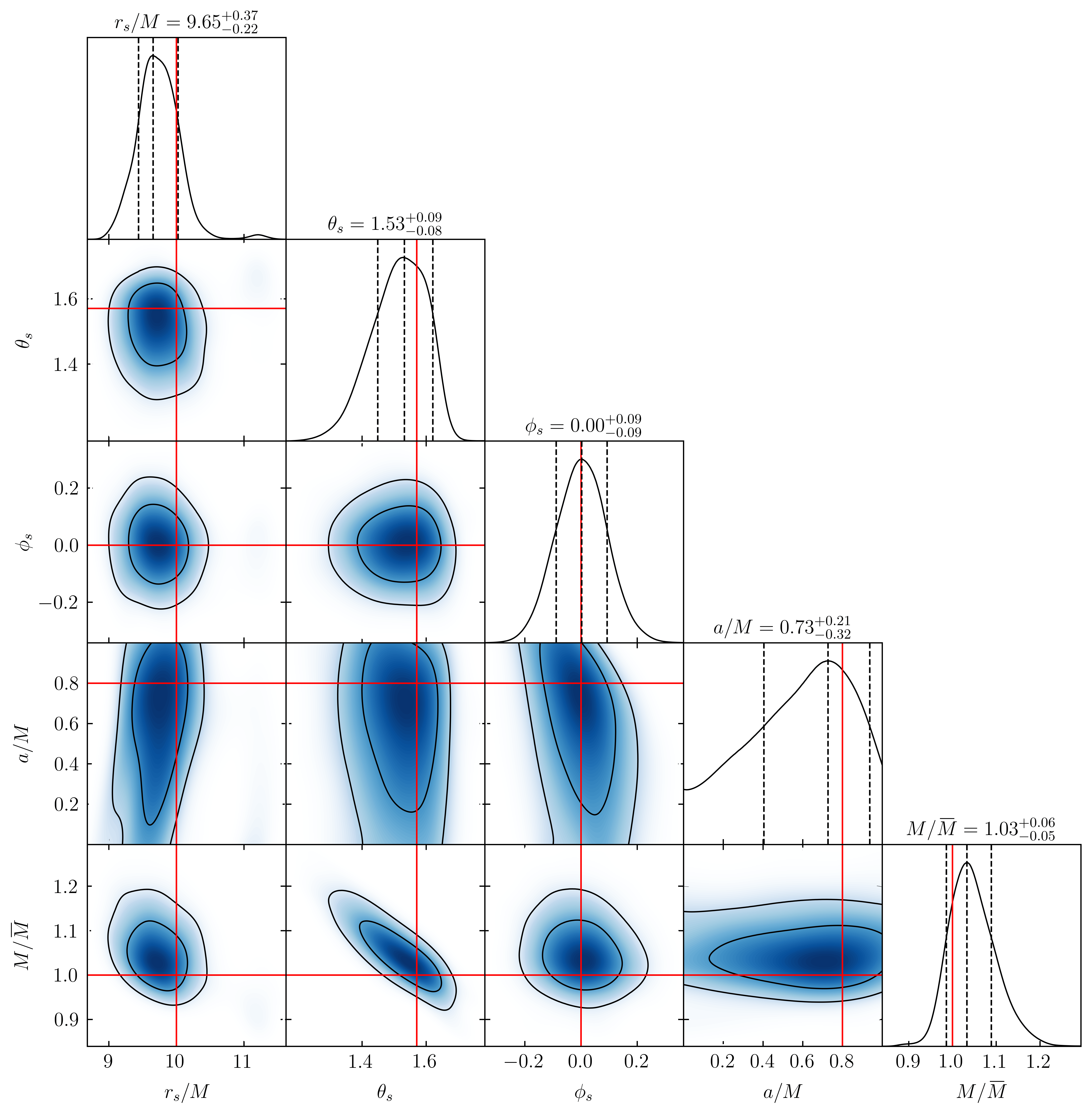}
    \caption{Posterior distributions for BH and hotspot location parameters derived from observing $N=0$ and $N=1$ photons from a hotspot outside M87$^*$, as detailed in Table~\ref{tab:set M87}. The spin projection PA and $\theta_o = {(163\pm 2)}^\circ$ are incorporated as predetermined priors. Red lines indicate the true values. On the diagonal, middle dashed lines mark the mode of the distributions, and the flanking dashed lines delineate the $1\sigma$ ($68.3\%$) credible interval. The determined values and their uncertainties are listed at the top of each plot. Contour lines in the off-diagonal plots represent the $1\sigma$ ($39.3\%$) and $2\sigma$ ($86.4\%$) credible regions.}
    \label{fig:mc_M87}
\end{figure*}

Figure~\ref{fig:mc_M87} presents the results of the MCMC simulation, visualized using the Python package \textit{GetDist}~\cite{Lewis:2019xzd}. As anticipated, the true values of the parameters fall within the $1\sigma$ credible intervals. This method estimates the BH mass and spin to within $5\%$ and $40\%$ accuracy, respectively, from a single observation without assumptions about the source location or inclination angles. Multiple independent observations could enhance this resolution proportionally to the square root of their number.

\begin{table}[!htbp]
\renewcommand\arraystretch{1.3} %% row spacing
\setlength\tabcolsep{2.5pt} %% column spacing
\begin{tabular}{cc|cccc}
\hline\hline
Parameter & Truth & & Observable & Measurement & Uncertainty \\ \hline
 $M$ & $\overbar{M}$  & & $\alpha_0$ & $-7.45\,\overbar{M}$ & $\overbar{M}$ \\ 
 $a/M$ & $0.8$ & & $\beta_0$ & $7.32\,\overbar{M}$ & $\overbar{M}$ \\ 
 $r_s/M$ & $10$  & & $\alpha_1$ & $1.62\,\overbar{M}$ & $2\,\overbar{M}$\\ 
 $\theta_s$ & $\ang{90}$  & & $\beta_1$ & $-5.30\,\overbar{M}$ & $2\,\overbar{M}$\\
 $\phi_s$ & $\ang{-45}$  & & $\Delta t$ & $29.57\,\overbar{M}$ & $0.5\,\overbar{M}$\\ \hline
\end{tabular}
\caption{\textit{Left}: Parameters to be determined for a hotspot outside M87$^*$ and their assumed true values; \textit{Right}: Observables for $N=0$ and $N=1$ photons emitted by the hotspot, along with their measurement values  (assumed to match theoretical true values) and measurement uncertainties. The spin projection PA and $\theta_o = {(163\pm 2)}^\circ$ are predetermined.}
\label{tab:set M87}
\end{table}

%%%%%%%%%%%%%%%%%%%%
\subsection{Sgr A$^*$ with Undetermined PA and $\theta_o$}
\label{subsec:SgrA}
%%%%%%%%%%%%%%%%%%%%

The absence of a jet observation for Sgr A$^*$ complicates the determination of its spin projection PA and inclination angle. Although recent observations by the GRAVITY collaboration suggest a preference for $\theta_o \sim \ang{157}$~\cite{GRAVITY:2023avo}, in this subsection we treat it as undetermined to demonstrate the capability of independent EHT/ngEHT measurements to infer these parameters through hotspot correlation. Unlike the scenario described in Sec.~\ref{subsec:M87}, a single hotspot observation providing $5$ constraints is insufficient to resolve all $7$ unknown parameters for Sgr A$^*$. At least two independent observations are necessary, yielding $10$ constraints, which matches the number of unknowns. Given the frequency of hotspot emergence around Sgr A$^*$—potentially daily—this requirement is practically feasible.

To accommodate the undetermined spin projection PA, we introduce a Cartesian coordinate system $(X,\,Y)$ on the image plane, which relates to the $(\alpha,\,\beta)$ coordinates through a rotation transformation
\begin{equation}
    \begin{pmatrix}
        X \\ Y
    \end{pmatrix} = \begin{pmatrix}
    \cos\PA & -\sin\PA \\ 
    \sin\PA & \cos\PA
    \end{pmatrix} \begin{pmatrix}
        \alpha \\ \beta
    \end{pmatrix}.
    \label{eq:XY}
\end{equation}
For each hotspot, the positions of the $N=0$ and $N=1$ images are denoted by $(X_0^{(j)},\,Y_0^{(j)})$ and $(X_1^{(j)},\,Y_1^{(j)})$ respectively, with a corresponding time delay $\Delta t^{(j)}$.

\begin{table}[t]
\renewcommand\arraystretch{1.3} %% row spacing
\setlength\tabcolsep{2.5pt} %% column spacing
\begin{tabular}{cc|cccc}
\hline\hline
Parameter & Truth & & Observable & Measurement & Uncertainty \\ \hline
 $M$ & $\overbar{M}$  & & $X_0^{(a)}$ & $-7.39\,\overbar{M}$ & $\overbar{M}$ \\ 
 $a/M$ & $0.8$ & & $Y_0^{(a)}$ & $6.99\,\overbar{M}$ & $\overbar{M}$ \\ 
 $\theta_o$ &  $\ang{157}$ & & $X_1^{(a)}$ & $1.53\,\overbar{M}$ &$2\,\overbar{M}$ \\ 
 $\PA$ & $0$  & & $Y_1^{(a)}$ & $-5.32\,\overbar{M}$ & $2\,\overbar{M}$\\ 
 $r_s^{(a)}/M$ & $10$  & & $\Delta t^{(a)}$ & $30.88\,\overbar{M}$ & $2\,\overbar{M}$\\
 $\theta_s^{(a)}$ & $\ang{90}$  & & $X_0^{(b)}$ & $7.34\,\overbar{M}$ & $\overbar{M}$\\ 
 $\phi_s^{(a)}$ & $\ang{-45}$  & & $Y_0^{(b)}$ & $-1.78\,\overbar{M}$ & $\overbar{M}$\\ 
 $r_s^{(b)}/M$ & $8$  & & $X_1^{(b)}$ & $-3.57\,\overbar{M}$ & $2\,\overbar{M}$\\ 
 $\theta_s^{(b)}$ & $\ang{120}$  & & $Y_1^{(b)}$ & $ 3.34\,\overbar{M}$ & $2\,\overbar{M}$\\ 
 $\phi_s^{(b)}$ & $\ang{90}$  & & $\Delta t^{(b)}$ & $ 28.42\,\overbar{M}$ & $2\,\overbar{M}$\\ \hline
\end{tabular}
\caption{\textit{Left}: Parameters to be determined for two hotspots outside Sgr A$^*$ and their assumed true values, including PA of spin projection and $\theta_o$. \textit{Right}: Observables for $N=0$ and $N=1$ photons from the hotspots, along with their true values and measurement uncertainties.
}
\label{tab:SgrA}
\end{table}

\begin{figure*}
    \centering
    \includegraphics[width=0.95\textwidth]{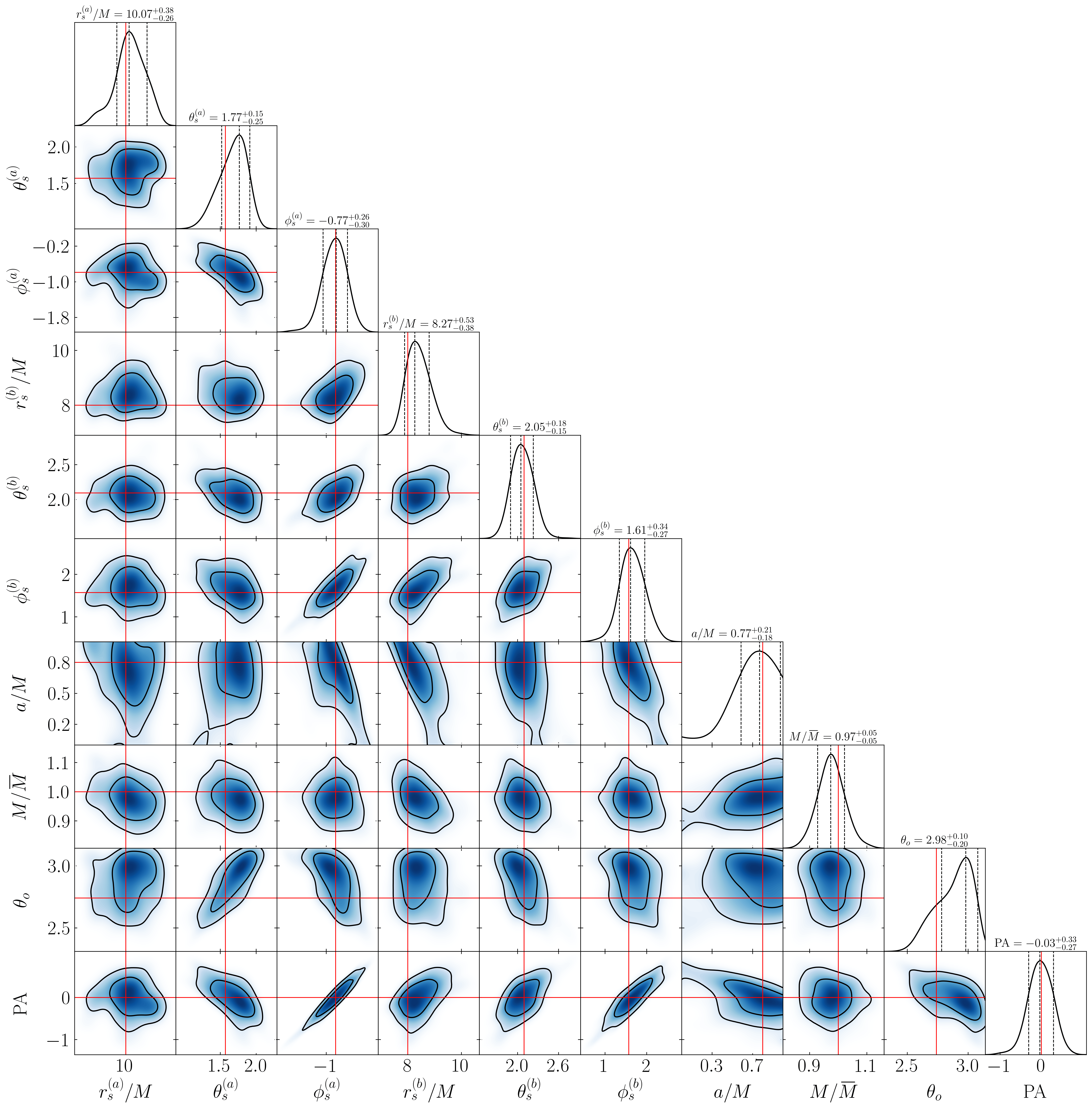}
    \caption{Posterior distributions for BH and hotspot location parameters obtained from observations of $N=0$ and $N=1$ photons from two hotspots outside Sgr A$^*$, which include spin projection PA and $\theta_o$, as detailed in Table~\ref{tab:SgrA}. The notation in this figure is consistent with that used in Fig.~\ref{fig:mc_M87}.}
    \label{fig:mcSgrA}
\end{figure*}

Table~\ref{tab:SgrA} displays the assumed true values for BH spacetime parameters and hotspot positions on the left side for two examples involving Sgr A$^*$. The right side lists the corresponding true values and measurement uncertainties for the observables. The MCMC results, illustrated in Fig.~\ref{fig:mcSgrA}, demonstrate that most parameters, such as BH mass and spin, are effectively constrained, showing posterior distributions with distinct central peaks; the true values fall within the $1\sigma$ credible intervals. However, accurately determining $\theta_o$ is more problematic, with its true value lying outside the $1\sigma$ interval. This challenge arises because both the position angle shift and the time delay between the $N=0$ and $N=1$ photons are only marginally influenced by $\theta_o$ when it is nearly face-on~\cite{Gralla:2019drh,Hadar:2020fda}. This degeneracy can be mitigated through multiple hotspot correlation observations. Similarly, the accuracy of measuring universal spacetime parameters, such as BH mass and spin, is expected to improve as the number of independent hotspot observations increases, which is promising for Sgr A$^*$.

%%%%%%%%%%%%%%%%%%%%%%%%%%%%%%%%%%%%%%%%%%%%%%%%%
\section{Discussion}\label{sec:dis}
%%%%%%%%%%%%%%%%%%%%%%%%%%%%%%%%%%%%%%%%%%%%%%%%%

Null geodesics connecting two spatial points in Kerr spacetime can exhibit multiple solutions, attributed to the strong gravitational lensing by the BH. This study develops a systematic framework to identify these solutions, categorizing them by the number of orbits executed around the BH. Each geodesic is uniquely determined by its conserved quantities in Kerr spacetime and the initial momentum signs.

A direct application of this methodology is to BH imaging, particularly in connecting a pointlike emission source outside a BH to a distant observer. This forward ray tracing approach proves more efficient than traditional backward ray tracing, especially in modeling hotspots and calculating higher-order images. By applying perturbative adjustments to the original geodesics around the emission point, our method can map a finite-size emission to distinct regions on the observer plane. 
Higher-order images, typically positioned near the critical curve, sequentially compress in the direction perpendicular to the curve while maintaining a stable amplification rate parallel to it. Consequently, the total area magnification rate exhibits exponential decay as the image level $N$ increases.

Our method is powerful in identifying lensed photons and conducting spacetime tomography, particularly for measuring BH mass and spin. Utilizing observations from the EHT and ngEHT that capture both direct and lensed images, our approach can precisely determine BH characteristics. For M87$^*$, with predetermined spin projection and inclination angle, a single measurement suffices to resolve the BH's mass, spin, and the emission location. In the case of Sgr A$^*$, two independent measurements of hotspot correlations are enough to determine all relevant parameters. Given the daily occurrence of flares from hotspots around Sgr A$^*$, the prospects for successful detection are promising, especially with ngEHT, which will offer tracking of hotspot motion~\cite{Emami:2022ydq} and multifrequency measurement capabilities~\cite{Chael:2022meh}. The accuracy in determining universal spacetime parameters will improve as more data accumulates.

Importantly, the forward ray tracing method extends beyond photons to gravitational waves, which also adhere to null geodesics in the geometric optics limit. For instance, gravitational waves emitted by stellar-mass BH binaries near supermassive BHs can generate echoes via direct and lensed geodesics~\cite{Chen:2017xbi,DOrazio:2019fbq,Gondan:2021fpr,Basak:2022fig,Savastano:2022jjv,Zhang:2024ibf,Kubota:2024zkv}. Supermassive BH binaries might also exhibit self-lensing signatures during inspiral phases~\cite{DOrazio:2017ssb, Ingram:2021gar, Kelley:2021yfc}. Our method can potentially predict modulations in gravitational wave signals from these events, thus unveiling their orbital dynamics.

Given that our approach utilizes conserved quantities in Kerr spacetime, an intriguing question arises: can forward ray tracing be adapted for perturbed spacetimes, such as those with metric perturbations~\cite{Giddings:2016btb, Giddings:2019jwy, Wang:2019skw, Chen:2022kzv, Zhu:2023omf,Zhong:2024ysg}? Additionally, exploring the interplay between forward ray tracing and radiative transfer could yield insights, particularly for the propagation of polarimetric signals~\cite{Chen:2019fsq, Himwich:2020msm, Chen:2021lvo, Chen:2022oad}.

\hspace{5mm}
%%%%%%%%%%%%%%%%%%%%%%%%%%%%%%%%%%

%%%%%%%%%%%%%%%%%%%%%%%%%%%%%%%%%%
\begin{acknowledgments}
%%%%%%%%%%%%%%%%%%%%%%%%%%%%%%%%%%
We are grateful to Xian Chen, Daniel D'Orazio, Shahar Hadar, Yosuke Mizuno, Daniel Palumbo, Diogo Ribeiro, Paul Tiede, Luka Vujeva, Bo Wang, George Wong, and Xiao Xue for useful discussions.
L.Z. acknowledges financial support from Peking University for his visit to the Niels Bohr Institute.
Z.Z. acknowledges financial support from China Scholarship Council (No.~202106040037).
The Center of Gravity is a Center of Excellence funded by the Danish National Research Foundation under grant No. 184. V.C. and Y. C. acknowledge support by VILLUM Foundation (grant no. VIL37766) and the DNRF Chair program (grant no. DNRF162) by the Danish National Research Foundation. V.C. is a Villum Investigator and a DNRF Chair.  V.C. acknowledges financial support provided under the European Union's H2020 ERC Advanced Grant “Black holes: gravitational engines of discovery” grant agreement no. Gravitas–101052587.
Views and opinions expressed are however those of the author only and do not necessarily reflect those of the European Union or the European Research Council. Neither the European Union nor the granting authority can be held responsible for them. This project has received funding from the European Union's Horizon 2020 research and innovation programme under the Marie Sklodowska-Curie grant agreement No 101007855 and No 101131233.

\end{acknowledgments}

%\bibliography{references.bib}


\end{document}